\pdfoutput=1
\documentclass[11pt,a4paper]{article}
\usepackage{jheppub}
\usepackage[utf8]{inputenc}
\usepackage{epsfig,amsfonts,amsthm}
\usepackage[normalem]{ulem}
\usepackage{amsmath,amssymb}
\usepackage{array}
\usepackage{amsmath}
\usepackage{amsfonts}
\usepackage{amssymb}
\usepackage{subfig}
\usepackage{wrapfig}
\usepackage{graphicx}
\usepackage{slashed}
\usepackage{soul}
\usepackage{color}
\usepackage{cancel}
\usepackage[english]{babel}
\usepackage{url}
\usepackage{placeins}

\newcommand{\be}{\begin{equation}}
\newcommand{\ee}{\end{equation}}
\newcommand{\bea}{\begin{eqnarray}}
\newcommand{\eea}{\end{eqnarray}}

\definecolor{grey}{cmyk}{0,0,0,0.75}
\definecolor{tangerine}{cmyk}{0,0.5,1,0}
\definecolor{darkgreen}{cmyk}{1,0,1,0.23}
\definecolor{Red}{rgb}{1,0,0}
\definecolor{Blue}{rgb}{0,0,1}
\definecolor{Green}{rgb}{0,1,0}
\definecolor{Grey}{cmyk}{0,0,0,0.75}
\definecolor{Tangerine}{cmyk}{0,0.5,1,0}
\definecolor{Darkgreen}{cmyk}{1,0,1,0.23}
\definecolor{Cyan}{cmyk}{1,0,0,0}
\definecolor{Yellow}{cmyk}{0,0,1,0}
\definecolor{darkblue}{cmyk}{1,0.69,0,0.11}

\def\lsim{\mathrel{\rlap{\lower4pt\hbox{\hskip1pt$\sim$}}
    \raise1pt\hbox{$<$}}}         
\def\gsim{\mathrel{\rlap{\lower4pt\hbox{\hskip1pt$\sim$}}
    \raise1pt\hbox{$>$}}}         

\newcommand{\Hz}{\,\mathrm{Hz}}

\newcommand{\vw}{v_w}
\newcommand{\snr}{\rho}

\def\D{\mathrm{d}}

\def\beq{\begin{equation}}
\def\eeq{\end{equation}}
\def\bea{\begin{eqnarray}}
\def\eea{\end{eqnarray}}

\def\<{\left\langle}
\def\>{\right\rangle}

\def\lsim{\mathrel{\rlap{\lower4pt\hbox{\hskip1pt$\sim$}}
    \raise1pt\hbox{$<$}}}         
\def\gsim{\mathrel{\rlap{\lower4pt\hbox{\hskip1pt$\sim$}}
    \raise1pt\hbox{$>$}}}         

\def\D{\mathrm{d}}

\def\beq{\begin{equation}}
\def\eeq{\end{equation}}
\def\bea{\begin{eqnarray}}
\def\eea{\end{eqnarray}}

\def\<{\left\langle}
\def\>{\right\rangle}

\newcommand{\bt}{\begin{tabular}}
\newcommand{\et}{\end{tabular}}

\newcommand{\mum}{\mu_{HS}}

\usepackage{hyperref}
\newcommand{\hc}[1]{#1^{\dagger}}

\newcommand{\abs}[1]{|#1|}

\begin{document}
\bibliographystyle{JHEP}

\preprint{HIP-2020-24/TH, LTH 1243}

\title{Pseudo-Goldstone dark matter: gravitational waves and direct-detection blind spots}

\author[a,b]{Tommi Alanne,}
\author[c]{Nico Benincasa,}
\author[d]{Matti Heikinheimo,}
\author[c]{Kristjan Kannike,}
\author[d]{\\Venus Keus,}
\author[c]{Niko Koivunen,}
\author[d]{Kimmo Tuominen}

\affiliation[a]{Department of Mathematical Sciences, University of Liverpool,\\
  Liverpool, L69 7ZL, United Kingdom}
\affiliation[b]{Max-Planck-Institut f\"{u}r Kernphysik,
  Saupfercheckweg 1, 69117 Heidelberg, Germany}
\affiliation[c]{National Institute of Chemical Physics and Biophysics,
    Ravala 10, Tallinn 10143, Estonia}
\affiliation[d]{Department of Physics and Helsinki Institute of Physics,\\
  Gustaf Hallstromin katu 2, FIN-00014 University of Helsinki,
  Finland}

\emailAdd{tommi.alanne@liverpool.ac.uk}
\emailAdd{nico.alexis.benincasa@ut.ee}
\emailAdd{matti.heikinheimo@helsinki.fi}
\emailAdd{kristjan.kannike@cern.ch}
\emailAdd{venus.keus@helsinki.fi}
\emailAdd{niko.koivunen@kbfi.ee}
\emailAdd{kimmo.i.tuominen@helsinki.fi}

\abstract{Pseudo-Goldstone dark matter is a thermal relic with momentum-suppressed
direct-detection cross section. We study the most general model of
pseudo-Goldstone dark matter arising from the complex-singlet extension of the
Standard Model. The new U(1) symmetry of the model is explicitly broken down to
a CP-like symmetry stabilising dark matter. We study the interplay of
direct-detection constraints with the strength of cosmic phase transitions and
possible gravitational-wave signals. While large U(1)-breaking interactions can generate a large direct-detection cross section, there are blind spots where the cross section is suppressed. We find that sizeable cubic couplings can
give rise to a first-order phase transition in the early universe.
We show that there exist regions of the parameter space where
the resulting gravitational-wave signal can be detected in future by the
proposed Big Bang Observer detector.}

\maketitle
\newpage

\section{Introduction}

Recent direct-detection results have imposed severe
constraints~\cite{Akerib:2016vxi,Aprile:2018dbl,Cui:2017nnn} on some of the most
popular dark matter (DM) frameworks, such as the real-scalar-singlet extension of the Standard Model (SM).
Pseudo-Goldstone DM is a simple framework with a naturally small direct-detection cross section due to suppressed scattering rates at low momentum
transfer resulting from the derivative interactions of the Goldstone boson.

In the minimal pseudo-Goldstone DM model \cite{Gross:2017dan,Huitu:2018gbc,Alanne:2018zjm,Azevedo:2018oxv,Karamitros:2019ewv,Cline:2019okt,Arina:2019tib} (see also refs.~\cite{Barger:2008jx,Chiang:2017nmu}),
the global U(1) symmetry is explicitly
broken down to a $\mathbb{Z}_{2}$ symmetry by the DM mass term.
In this case, the electroweak phase transition is of second
order~\cite{Kannike:2019wsn}. If the U(1) symmetry is explicitly broken to $\mathbb{Z}_{3}$ \cite{Kannike:2019mzk} or to nothing, the resulting cubic
terms of the general model can induce, in part of the parameter space, strong first-order phase transitions \cite{Witten:1984rs,Hogan:1984hx,Steinhardt:1981ct}
leading to a gravitational-wave (GW) signal detectable by LISA~\cite{Audley:2017drz,Baker:2019nia}, DECIGO~\cite{Seto:2001qf,Kawamura:2006up,Yagi:2011wg,Isoyama:2018rjb}, or BBO~\cite{Crowder:2005nr,Corbin:2005ny,Harry:2006fi}. These cubic couplings have also been considered in a different context,
e.g., in refs.~\cite{Jiang:2015cwa,Alves:2018oct,Alves:2018jsw}.

The tree-level direct-detection cross section vanishes\,---\,in the
limit of zero momentum transfer\,---\,only in the case of the $\mathbb{Z}_{2}$-
symmetric pseudo-Goldstone DM model
with the U(1) symmetry softly broken by a mass term \cite{Gross:2017dan}.
In general, interaction
terms that explicitly break the U(1) also yield contributions to the
direct-detection cross section at tree level which do not vanish at zero
momentum transfer. At one-loop level the zero-momentum-transfer cross section
is non-vanishing already in the $\mathbb{Z}_{2}$-symmetric model \cite{Alanne:2018zjm,Azevedo:2018exj,Ishiwata:2018sdi}.

These features can be understood as follows. The vanishing of the cross section
at zero momentum transfer is a manifestation of the underlying continuous
global symmetry; in the absence of any explicit symmetry-breaking terms, the
prospective DM candidate is an exact Goldstone boson and therefore has only
derivative interactions, which yield zero cross
section in the $t\to 0$ limit in elastic scattering processes. An exact
Goldstone boson would of course be massless, which is why the minimal model must
contain at least the $\mathbb{Z}_{2}$-symmetric mass term which breaks U(1),
but yields a vanishing zero-momentum-transfer cross section at tree level.
Any other operator breaking the symmetry explicitly, and thereby contributing
to the mass of the pseudo-Goldstone boson, results in a non-vanishing
zero-momentum-transfer cross section already at tree level.

Our aim is to study the most general model of pseudo-Goldstone DM arising from
the complex-singlet extension of the SM; the parameter space studied
here has some overlap with that of ref.~\cite{Chao:2017vrq}. The scalar potential in this model is
inevitably CP-conserving \cite{Branco:1999fs}, and the only symmetry of the
potential is the CP-like $S \to S^{*}$ invariance which stabilises the
imaginary part of the complex singlet, $S$. In particular,
we seek to study the correlation
between the direct-detection cross section and the strength of the electroweak
phase transition. We take into account theoretical constraints from
perturbativity, unitarity and vacuum stability together with experimental
constraints on the invisible width of the Higgs boson and on the Higgs-singlet
mixing angle.

Our main result relevant for direct detection of pseudo-Goldstone DM is
that in this model certain combinations of 
parameters appear in both the pseudo-Goldstone mass
and the $t \to 0$ cross section. Setting such a combination to zero
then makes the direct-detection cross section vanish at tree level (but not
at loop level) on a slice of the parameter space. On the other hand,
we will uncover regions of parameter space where the model has sufficiently
strong first-order finite-temperature phase transition so that the
associated gravitational-wave signal can be detected in future satellite
observatories. There is a significant correlation between the strength
of the gravitational-wave signal and direct-detection cross section: Increasing
the former will also make the pseudo-Goldstone DM more easily detectable, while suppressing the latter typically results in a weak phase transition.

The paper is organised as follows: We introduce the general complex-singlet
model in section~\ref{sec:model}. In section~\ref{sec:cross} we discuss
DM phenomenology including direct and indirect detection.
Cosmic phase
transitions are considered in section~\ref{sec:pt}, and we conclude in
section~\ref{sec:conclusions}. The appendices contain some more technical details:
In appendix~\ref{sec:shift} we give the formulae for shifting away the linear term
in the potential, appendix~\ref{sec:x:sections} lists the annihilation cross sections,
appendix~\ref{app:DD1loop} shows how the one-loop contribution to the direct-detection
cross section is modified in the presence of cubic symmetry-breaking terms, and in
appendix~\ref{sec:rges} we list the one-loop renormalisation group equations for the model.

\section{General complex-singlet model}
\label{sec:model}

We consider a model where the scalar sector consists
of the SM Higgs doublet, $H$, and a complex singlet, $S$. The model is by construction CP-conserving \cite{Branco:1999fs}, i.e. invariant under the CP-like transformation $S\rightarrow S^{\ast}$.
We write the potential as
\begin{equation}
  V_{\rm tree} = V_{0} + V_{\rm br},
\label{original_potential}
\end{equation}
where
\begin{equation}
  V_{0} = \mu_{H}^{2} \hc{H} H + \mu_{S}^{2} S^{*} S
  + \lambda_{H} (\hc{H} H)^{2}
  + \lambda_{HS} (\hc{H} H) S^{*} S + \lambda_{S} (S^{*} S)^{2}
\end{equation}
is invariant under a global U(1) transformation $S\to e^{i\phi}S$, while
the remaining part explicitly breaks the U(1) symmetry:
\begin{equation}
\begin{split}
  V_{\rm br} =& \frac{1}{\sqrt{2}} \mu_{1}^{3} (S + S^{*})
    +\frac{1}{2} \mu_{S}^{\prime 2} (S^{2} + S^{* 2})\\
  &+ \frac{1}{2 \sqrt{2}} \mum \hc{H} H (S + S^{*})
    + \frac{1}{2} \mu_{3} (S^{3} + S^{* 3})
    + \frac{1}{2} \mu_{3}^{\prime} S S^{*} (S + S^{*})\\
  &+ \frac{1}{2} \lambda_{HS}^{\prime} \hc{H} H (S^{2} + S^{* 2})
  + \frac{1}{2} \lambda_{S}^{\prime} (S^{4} + S^{* 4})
    + \frac{1}{2} \lambda_{S}^{\prime\prime} S S^{*} (S^{2} +S^{* 2})\,.
\end{split}
\label{VU1Br}
\end{equation}
The minimal $\mathbb{Z}_{2}$-symmetric pseudo-Goldstone DM model contains only the U(1)-symmetric potential and the explicit symmetry-breaking $\mu_{S}^{\prime 2}$ mass term; the potential \eqref{original_potential} is the most general setup.
In the unitary gauge the scalar multiplets are parametrised as
\be
H=\frac{1}{\sqrt{2}}\left(\begin{array}{c}
0\\
v+h
\end{array}
\right),\quad \textrm{and} \quad S=\frac{1}{\sqrt{2}}(w+s+i\chi),
\ee
where $v=246.22\ {\rm GeV}$ is the usual electroweak vacuum expectation value (vev), and $w$ is the vev of the singlet scalar. The $\lambda_{S}^{\prime\prime}$ term
produces an independent contribution only to the $\chi^4$ vertex; elsewhere it
can be absorbed by a redefinition of the other couplings. Consequently,
we will set $\lambda_S^{\prime\prime}=0$ in what follows.

Minimising the potential,
the mass of the CP-odd field, $\chi$, is calculated to be
\be
  m_\chi^2 = -2 \mu_S^{\prime\,2}-\lambda'_{HS}v^2-4\lambda'_S
  w^2-\frac{1}{2\sqrt{2}}(9\mu_3+\mu'_3)w-\frac{1}{4}\mum\frac{v^2}{w}-\frac{\mu_1^3}{w}.
\label{chi_massSq}
\ee
The real part of the singlet, $s$, mixes with the neutral component of the doublet, $h$, resulting in two mass eigenstates, $h_1$ and $h_2$.
The particle ${h_1=h \cos\theta - s \sin\theta}$ is identified as the SM-like Higgs boson with mass $m_{1}=125.1$ GeV,
and the orthogonal linear combination $h_2$ is another
CP-even scalar with mass $m_{2}$.
The mixing angle $\theta$ is given by
\be
 \tan 2\theta=-\frac{8(\lambda_{HS}+\lambda'_{HS})v w^2
+4\mum vw}{8\lambda_H v^2w-8(\lambda_S+\lambda'_S
)w^3 + 3\sqrt{2}(\mu_3+\mu'_3)w^2+\mum v^2
+4\mu_1^3}.
\ee
We replace the parameters $\mu_H^2$, $\mu_S^2$, $\mu_S^{\prime 2}$, $\lambda_H$,
$\lambda_S$ and $\lambda_{HS}$ appearing in the potential with physical
parameters $m_{1}$, $m_{2}$, $m_\chi$, $\theta$, $v$ and $w$.
Fixing the value of the electroweak vev and the mass of the Higgs boson
to the known values reduces the number of independent parameters by two.
The independent parameters are then further constrained
by collider searches and cosmological observations.
Let us briefly discuss the phenomenological constraints which are
relevant for our model.

\paragraph{Stability of the potential and unitarity.}

To guarantee a stable vacuum, the potential has to be bounded from below.
This is in particular relevant for
large field values, where we can neglect dimensionful terms.
Imposing the co-positivity condition \cite{Kannike:2012pe,Kannike:2016fmd} on the matrix of quartic couplings requires
\bea
\label{eq:vacuum:stability}
&&
\lambda_{H} > 0,   \qquad
\lambda_{S} - |\lambda'_{S}| > 0,
\qquad
\lambda_{HS} - |\lambda'_{HS}|
+ 2 \sqrt{\lambda_{H}(\lambda_{S} + \lambda'_{S})} > 0,
\\
&&
4 (\lambda_{S} - \lambda'_{S})\sqrt{\lambda_{H}}
+ 2(\lambda_{HS} - \lambda'_{HS}) \sqrt{\lambda_{S} + \lambda'_{S}}
+ \sqrt{\lambda_{HR} \lambda_{HI} (\lambda_S-\lambda'_S)} > 0,
\nonumber\\
&&
~~~~\mbox{where} \hspace{4mm}
\lambda_{HR}\equiv \lambda_{HS} + \lambda'_{HS}
  + 2 \sqrt{\lambda_{H}(\lambda_{S} + \lambda'_{S})} > 0, \nonumber\\
&&
~~~~\hspace{14mm}
\lambda_{HI}\equiv \lambda_{HS} - \lambda'_{HS}
  + 2 \sqrt{\lambda_{H}(\lambda_{S} + \lambda'_{S})} > 0.
\nonumber
\eea

In addition to this, we  ensure that the point
\be
\langle H \rangle =\frac{1}{\sqrt{2}}(0\ \, v)^T,\qquad
\langle S \rangle =\frac{w}{\sqrt{2}},
\label{eq:vacuum}
\ee
is the global minimum of the potential.
Unitarity of the $S$ matrix for two-to-two elastic scattering constrains the
values of combinations of $\lambda$-parameters in the potential
at asymptotically large center-of-mass energies
\cite{Kanemura:1993hm, Akeroyd:2000wc
}.
In our numerical analysis with the \texttt{SARAH} package \cite{Goodsell:2018tti}, we determine the eigenvalues $\Lambda_i$
of the scattering matrix, require $\vert\Lambda_i\vert \le 1/2$ and implement the
resulting constraints on the quartic couplings.
In our model this condition translates into
\be
\vert\lambda_H\vert\leq 4\pi,~~\vert\lambda_{HS} + \lambda'_{HS}\vert\leq 8\pi,~~\vert\lambda_{HS} - \lambda'_{HS}\vert\leq 8\pi,~~\vert\lambda_{S} - 3 \lambda'_{S}\vert\leq 4\pi.
\ee
The three remaining eigenvalues $\Lambda_{1,2,3}$ of the scattering matrix are the solutions to a cubic equation
and lie in the interval \cite{10.2307/2285901,Belanger:2014bga}
\begin{align}
\label{unitarity}
\vert\Lambda_{1,2,3}\vert\leq & -2 \lambda_H - 4 \lambda_{S} - 4 \lambda'_{S} +
4 \sqrt{\lambda_H^2 +
2/3 (2 \lambda_{S}^2 + \lambda_{HS}^{2} + \lambda_{HS}^{\prime 2} +
6 \lambda_{S}^{\prime 2}) }.
 \end{align}

\paragraph{Constraints from collider experiments.}

From the two CP-even mass eigenstates, $h_1$ and $h_2$, we identify $h_1$ as the SM-like boson whose couplings are scaled by $\cos\theta$ with respect to the Higgs boson in the SM $h_{\text{SM}}$.
The signal strength of the decay of the SM-like Higgs boson
$h_1$ to final state $XX$ is defined as \cite{Khachatryan:2016vau}
\be
\mu_{XX}^{} \equiv
\frac{\sigma(gg\to h_1)}{\sigma(gg\to h_{\text{SM}} )}\times \frac{\text{Br}(h_1\to XX)}{\text{Br}(h_{\text{SM}}\to XX)}
\ee
and constrains the parameter space in each decay channel.
In particular, the signal strengths constrain the value of the mixing
angle to satisfy $\cos^2\theta\ge 0.9$~\cite{Beacham:2019nyx}.
The latest measurement of the width of an SM-like Higgs boson
gives $\Gamma^{h_1}_\text{tot}\,=3.2^{+2.8}_{-2.2}$ MeV, with 95\% CL limit on
$\Gamma_\text{tot} \leq 9.16$ MeV \cite{Sirunyan:2019twz}.
In our model, the total width of the SM-like Higgs boson can be modified\,---\,if kinematically allowed\,---\,through new decay channels $h_1\to\chi\chi$ and $h_1\to h_2h_2$\be
\textrm{Br}(h_1 \to XX) = \frac{\Gamma(h_1\to \chi \chi)+\Gamma(h_1
\to h_2h_2)}{\Gamma_{h_1}^{\rm SM}+\Gamma(h_1 \to \chi\chi)+\Gamma(h_1
\to h_2h_2)},
\label{invs1}
\ee
where the partial decay width for a new decay channel $h_1 \to XX$ is
\be
\Gamma(h_1\to XX)=\frac{g_{h_1XX}^{2}v^2}{32\pi m_{1}}\left(1-\frac{4m_{X}^{2}}{m_{1}^{2}}\right)^{1/2}.
\label{invdec}
\ee

Current experimental values provided by the ATLAS
and CMS experiments \cite{Khachatryan:2016whc,ATLAS-CONF-2018-031}  on the invisible branching ratio are
\be
\textrm{Br}(h_1 \to {\rm inv.}) <0.23-0.36, \label{invlimit}
\ee
where $h_1 \to {\rm inv.}$ represents the SM-like Higgs decay to the
DM candidate, $\chi$.
We will use the conservative limit $\textrm{Br}(h_1 \to {\rm inv.}) <0.23$ in the
following analysis.

\paragraph{Relic density measurements.}

The obsevations by the Planck satellite \cite{Aghanim:2018eyx}
show the abundance of DM to be
\be
\Omega_c\,h^2\,=\,0.120\,\pm\,0.001,  \label{PLANCK_lim}
\ee
where $h$ is the dimensionless Hubble parameter,
$H = 100\,h\, \textrm{km}/\textrm{s}/\textrm{Mpc}$.

In terms of the relative DM
abundance, $f_{\rm{rel}}=\Omega_\chi h^2/(\Omega_c h^2)$, we take
$f_{\rm{rel}}=1$ meaning that $\chi$ constitutes all of the
expected DM relic density.

Taking into account these preliminary constraints,
we then consider direct and indirect detection
of the DM candidate in our model to identify
regions of the parameter space surviving all the above constraints.

\section{Direct and indirect detection}
\label{sec:cross}


Let us now consider the cross sections relevant for the phenomenology of the
model. Our focus will be on understanding the suppression of the
direct-detection cross section, and how the situation changes in the presence of
various symmetry-breaking terms.

\subsection{Tree-level cross section}

Let us first review the suppression of the direct-detection cross section
at tree level. The CP-even scalar mass eigenstates couple to $\chi$ as
\be
\mathcal{L}=-\lambda_{h_1\chi\chi}\chi^2 h_1-\lambda_{ h_2\chi\chi}\chi^2 h_2.
\ee
Here the couplings are
\begin{equation}
\lambda_{h_1 \chi\chi} =
-\frac{1}{8w^2}\left(A(m_{1})\sin\theta+B\cos\theta\right), \quad
\lambda_{h_2\chi\chi} = \frac{1}{8w^2}\left(A(m_{2})\cos\theta-B\sin\theta\right),
\label{hDMcoupling}
\end{equation}
and we have defined
\be
A(m)=4m^2 w-32w^3 \lambda'_S+4\mu_1^3-\sqrt{2}(9\mu_3+\mu'_3)w^2+\mum v^2,
\ee
and
\begin{equation}
  B=2 v w (4\lambda'_{HS}w+\mum).
\end{equation}

The two  CP-even scalar mass eigenstates couple to the nucleon, $N$, via
the Higgs-boson Yukawa couplings as
\be
\mathcal{L}=-Y\cos\theta\, \bar{N}N h_1-Y\sin\theta\, \bar{N}N h_2,
\ee
where $Y=f_N m_N/v$, $m_N=0.946$~GeV is the nucleon mass, and we use $f_N = 0.3$ \cite{Alarcon:2011zs, Alarcon:2012nr, Cline:2013gha} for the effective
Higgs--nucleon coupling.

The spin-independent direct-detection cross section is given by

\be\label{lambda eff definition}
\frac{\D \sigma_{\rm SI}}{\D \Omega}=\frac{\lambda_{\rm SI}^2 f_N^2 m_N^2}{16
  \pi^2 m_\chi^2}\left(\frac{m_\chi m_N}{m_\chi+m_N}\right)^2,
\ee
where we have defined the effective DM--nucleon coupling $\lambda_{\rm SI}$ as
\be
\lambda_{\rm SI}^2 \equiv
\frac{1}{4 f_N^2 m_N^4}\lvert \mathcal{M}\rvert^2=
\frac{1}{m_N^2 v^2}\left[\frac{\lambda_{\chi\chi h_1}\cos\theta}{t-m_{1}^2}+\frac{\lambda_{\chi\chi h_2}\sin\theta}{t-m_{2}^2}\right]^2
(4m_N^2-t).
\ee
Writing the scalar couplings explicitly, cf. eq.~\eqref{hDMcoupling},
the effective direct-detection coupling in the $t\to 0$ limit becomes
\begin{align}
 \lambda_{\rm SI}= & \frac{m_{1}^2+m_{2}^2}{8\, v\, m_{1}^2 m_{2}^2}\Bigg\{\left(\frac{m_{2}^2-m_{1}^2}{m_{1}^2+m_{2}^2}\right)\bigg[\left(4\frac{\mu_1^3}{w^2}+\mum\frac{v^2}{w^2}-\sqrt{2}(9\mu_3+\mu'_3)-32w\lambda'_S\right)\sin 2\theta\nonumber\\
& \qquad\qquad\qquad\qquad\quad\ -2\frac{v}{w}(4w\lambda'_{HS}+\mum)\cos 2\theta\bigg] -2\frac{v}{w}(4w\lambda'_{HS}+\mum)\Bigg\}.
\label{DD coupling Niko}
\end{align}

As discussed in the introduction, the tree-level direct-detection cross section
vanishes in the $t\to0$ limit for U(1)-invariant interactions or
if only the $\mathbb{Z}_{2}$-symmetric $\mu_{S}^{\prime 2}$ mass term
is included. As explicitly shown by the above equation,
this is no longer true if any other symmetry-breaking interaction terms in the
potential, eq.~\eqref{VU1Br}, are present.

\begin{figure}[t!]
\begin{center}
\includegraphics[scale=0.57]{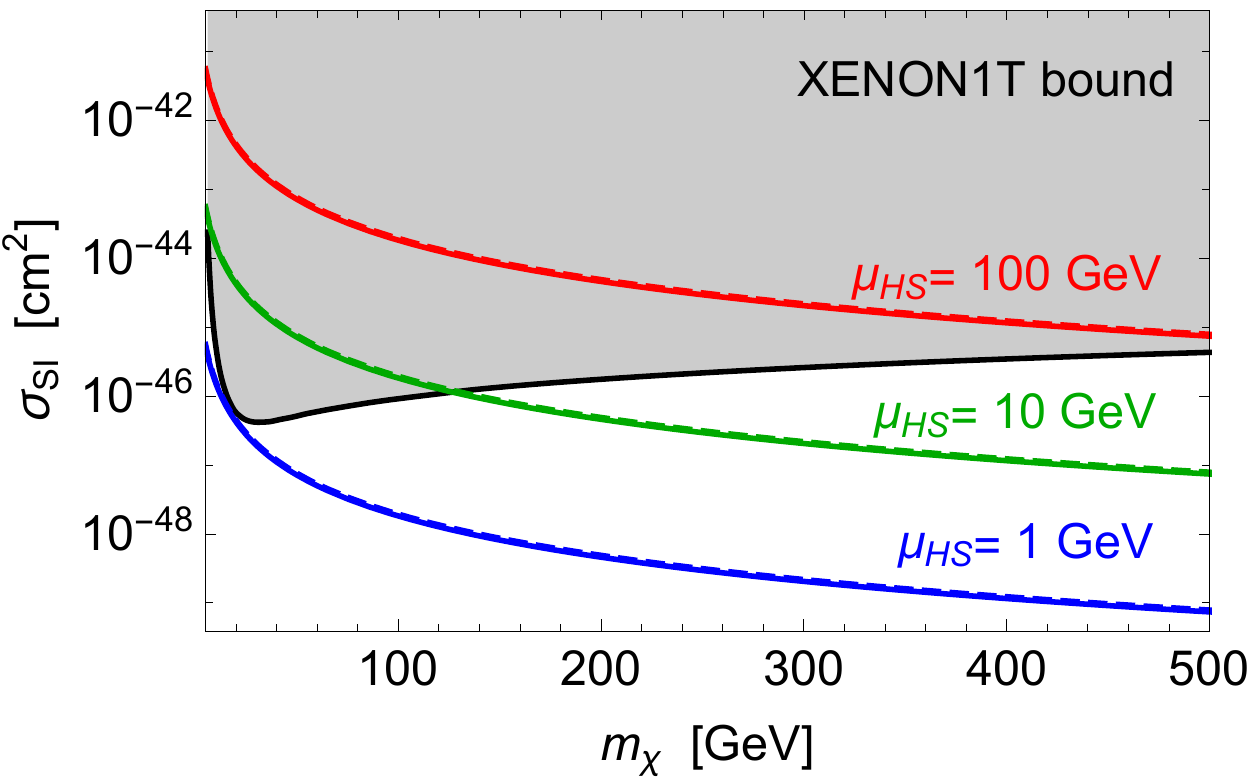}~~
\includegraphics[scale=0.57]{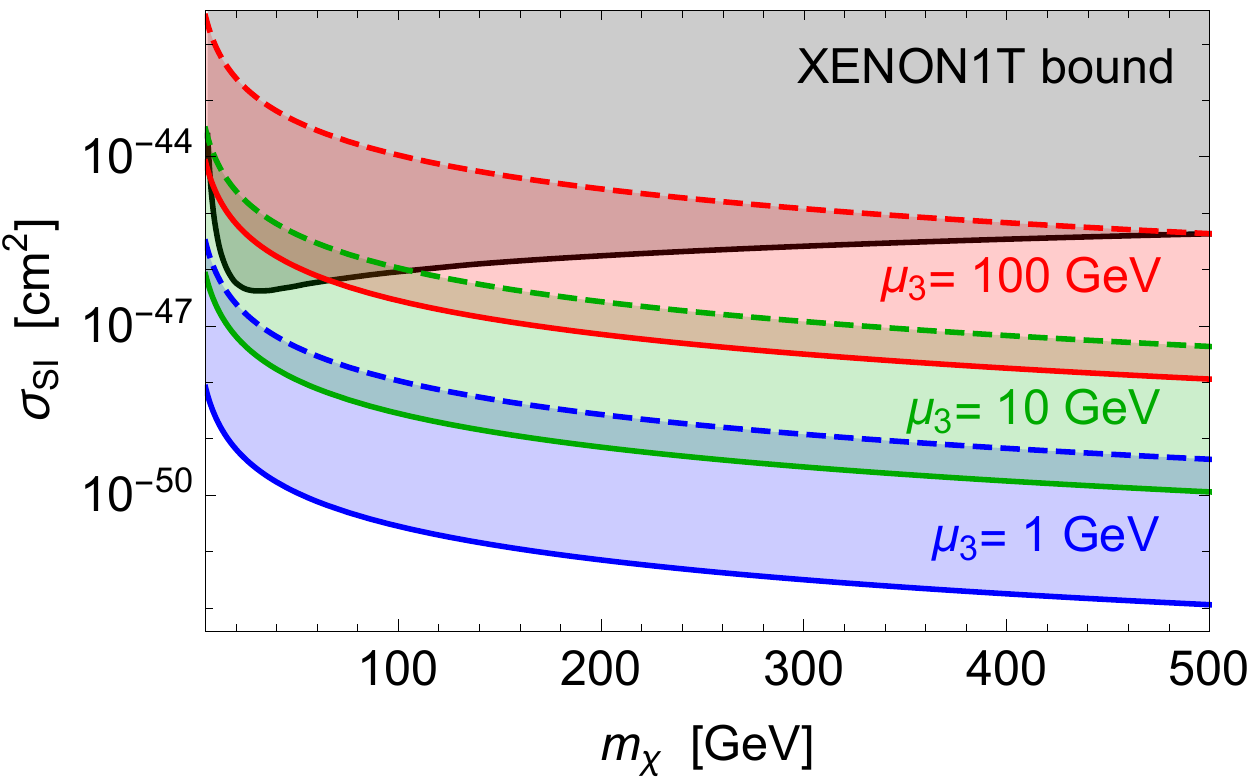}
\caption{The spin-independent DM--nucleus interaction cross section for different parameter values regardless of the relic density of the DM candidate, $\chi$. The cross-section values constrained by the XENON1T experiment~\cite{Akerib:2016vxi,Aprile:2018dbl} are shown in gray. In each plot, the solid lines represent the cross section for the minimum considered value for $\sin\theta = 0.01$ and the dashed lines represent the maximum considered value of $\sin\theta=0.2$. In all plots, $w=250$~GeV, $m_2=200$~GeV and the U(1)-breaking parameters which are not shown are set to zero, except for $\mu_{S}^{\prime 2}$.}
\label{DDLplots}
\end{center}
\end{figure}

To extract the effect of the symmetry-breaking terms on the direct-detection cross section, it is instructive to study the DM--nucleus interaction regardless of the relic-density contribution of the DM candidate, $\chi$. In figure \ref{DDLplots}, we
show the spin-independent DM--nucleus interaction cross section for the minimal $\mathbb{Z}_{2}$-symmetric model enhanced with only
one non-zero symmetry-breaking term in each plot.

\begin{figure}[h!]
\begin{center}
\includegraphics[scale=0.57]{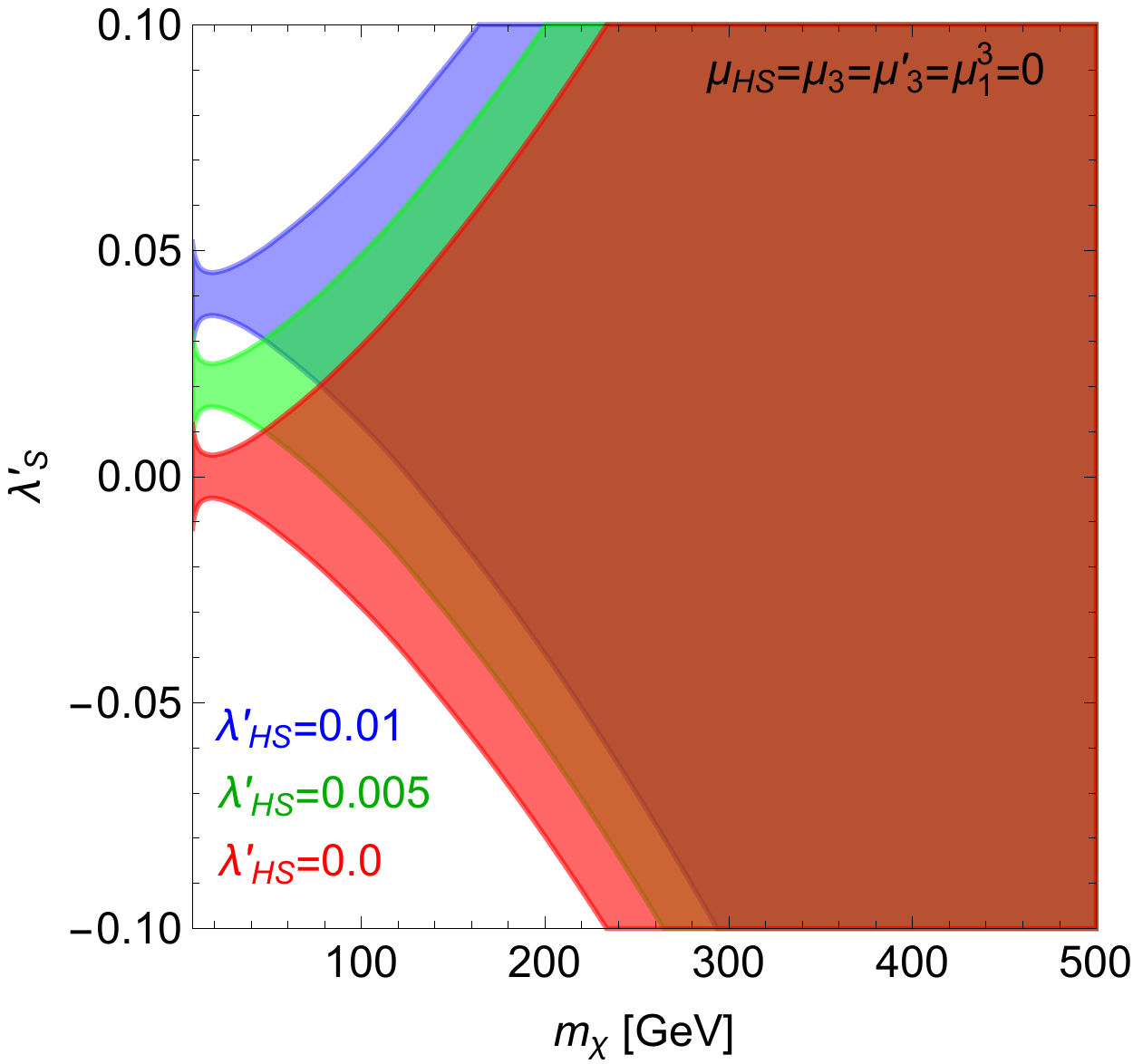}~~~
\includegraphics[scale=0.57]{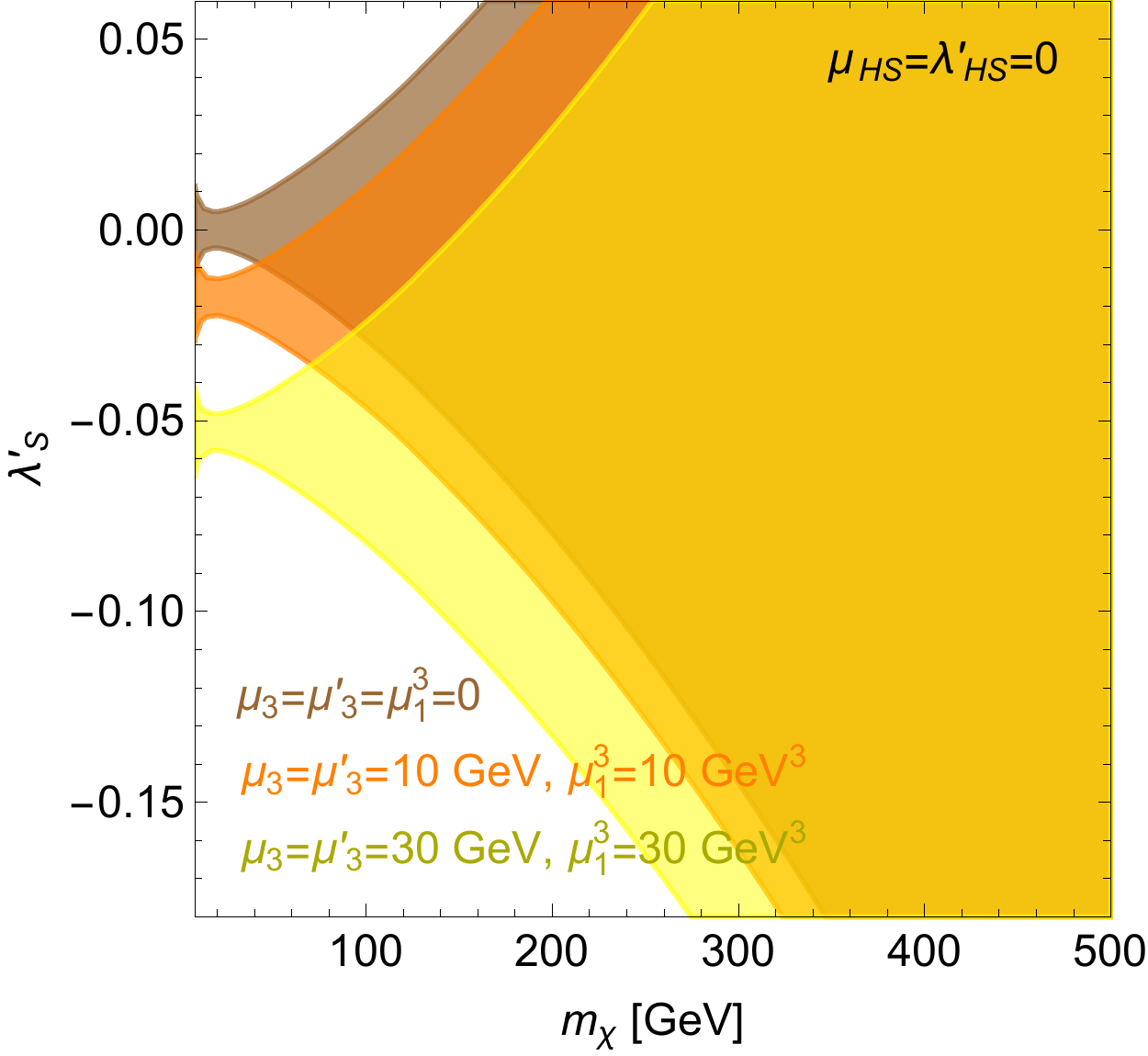}
\caption{Coloured regions are what is allowed by direct-detection bounds from XENON1T for $\sin\theta=0.1$, regardless of the relic-density contribution of the DM candidate. In the symmetry-breaking
potential all terms with odd powers of the singlet have been set to zero in the left panel, and in the right panel all doublet--singlet mixing terms have been set to zero. In both panels $w=250$ GeV and $m_2=200$ GeV.}
\label{DDgenplots}
\end{center}
\end{figure}

Figure \ref{DDgenplots} shows regions allowed by the XENON1T bound~~\cite{Akerib:2016vxi,Aprile:2018dbl} regardless of the relic-density contribution of $\chi$ for typical values of $w=250$ GeV, $m_2=200$ GeV and $\sin\theta=0.1$. In the left panel, all terms with odd powers of the singlet $S$ in the symmetry breaking
potential in eq.~\eqref{VU1Br} have been set to zero, i.e. $\mu_{HS}=\mu_3=\mu'_3=\mu_1 = 0$, and in the right panel all doublet-singlet mixing terms in eq.~\eqref{VU1Br} have been set to zero, i.e. $\mu_{HS}=0$ and $\lambda'_{HS}=0$.

\subsection{Cancellation regions}

Let us study the general formula for
the effective direct-detection coupling in the $t\to0$ limit in
eq.~\eqref{DD coupling Niko} in more detail.
As noted before, this result is generally non-zero in the presence of any of the
symmetry-breaking interactions in eq.~\eqref{VU1Br}. However, there are specific
combinations of the symmetry-breaking parameters which lead to a suppressed
$\lambda_{\rm SI}$ in the $t\to 0$ limit, thereby mimicking
the behaviour of the minimal $\mathbb{Z}_{2}$-symmetric model. We shall now
explore these cancellation conditions more closely.

Recall first, that in the non-linear representation, if only the U(1)-breaking $\mu_{S}^{\prime 2}$ mass term in eq.~\eqref{VU1Br} is present, we have
\begin{equation}
V_{\mathrm{int}}
\supset
-\frac{m_\chi^2}{w}\left[ \sin\theta\; h_1 - \cos\theta\; h_2 \right] \chi^2
 +
\frac{1}{w} \left[ \sin\theta \; h_1 - \cos\theta \; h_2 \right]
(\partial_\mu \chi)^2,
\end{equation}
yielding the following effective coupling for direct detection, where $p_1$ and $k_1$ are the momenta of the incoming and outgoing $\chi$-particles, resp.,
\begin{align}
\lambda_{ \mathrm{eff}}^{\mathrm{non-lin}}=&
\frac{\sin(2\theta)(m_1^2-m_2^2)}{vw (m_1^2-t)(m_2^2-t)}\left(-m_{\chi}^2
-(-p_1\cdot k_1)\right)\nonumber\\
=&\frac{\sin(2\theta)(m_1^2-m_2^2)}{vw (m_1^2-t)(m_2^2-t)}\left(-m_{\chi}^2-\frac{1}{2}(t-2m_{\chi}^2)\right)\nonumber\\
=&\ \frac{\sin(2\theta)(m_1^2-m_2^2)}{vw (m_1^2-t)(m_2^2-t)} \left(\frac{-t}{2}\right),
\label{eq:symmetricDD}
\end{align}
which explicitly shows that the direct-detection cross section vanishes in the $t\to 0$ limit.

Let us now study how the cubics $\mu_{3}$ and $\mu'_{3}$ affect the cross
section and pseudo-Goldstone mass. For simplicity, we take here
$\lambda'_S=0$ and $\mu^3_1=0$ and also set all
symmetry-breaking terms involving the Higgs boson to zero.
Representing the singlet field, $S$, as
\begin{equation}
  S= \frac{s+w}{\sqrt{2}} e^{i\chi/w},
\end{equation}
the cubic terms in the potential \eqref{original_potential}
can be written as
\begin{equation}
\label{eq:V3}
V_3 =\frac{1}{2 \sqrt{2}} (s+w)^3\left[\mu_3\cos\left(\frac{3\chi}{w}\right)
  +\mu_3^{\prime}\cos\left(\frac{\chi}{w}\right)\right].
\end{equation}
Minimisation of the potential yields
\begin{equation}
\mu_S^2 =-\lambda_Sw^2-\frac{3}{2\sqrt{2}}(\mu_3 + \mu'_3)w-\mu_S^{\prime\,2},
\end{equation}
and the mass of $\chi$ is calculated to be
\begin{equation}
  m_{\chi}^2=-2\mu_S^{\prime\,2}- \frac{1}{2 \sqrt{2}} (9\mu_3+\mu_3^{\prime})w.
  \label{eq:mchi:simpl}
\end{equation}
Let us now try to understand the origin of the contributions to the direct-detection cross section by explicitly relating the derivation of the $\mathbb{Z}_2$-symmetric case, eq.~\eqref{eq:symmetricDD}.
The relevant interaction terms in our simplified case are given by
\begin{equation}
  \label{eq:Vint}
  \begin{split}
    V_{\rm int}\supset& -\frac{1}{8w^2}\left[8\mu_{S}^{\prime 2}(s+w)^2+\sqrt{2}(9\mu_3
    +\mu_3^{\prime})(s+w)^3\right]\chi^2\\
    =&-\left[\mu_{S}^{\prime 2}+\frac{1}{4\sqrt{2}}(9\mu_3
    +\mu_3^{\prime})w\right]\chi^2
    -\frac{1}{w}\left[2\mu_{S}^{\prime 2}+\frac{3}{4\sqrt{2}}(9\mu_3
    +\mu_3^{\prime})w\right]s\chi^2+\dots\\
    =&\frac{1}{2}m_{\chi}^2\chi^2
    +\frac{1}{w}\left[m_{\chi}^{2}-\frac{1}{4\sqrt{2}}(9\mu_3
    +\mu_3^{\prime})w\right]s\chi^2+\dots,
  \end{split}
\end{equation}
where we used eq. \eqref{chi_massSq} to obtain the last equality.
The second term in the parenthesis of the last line of eq.~\eqref{eq:Vint}
does not cancel by the mass coming from the derivative terms, and thus there is
a contribution to direct detection from the cubic terms.

However, note how the mass $m_\chi$, given by eq.~\eqref{eq:mchi:simpl} in the simplified case at hand, is proportional to the same combination of parameters as in the direct-detection cross section. This shows explicitly that also in this case the suppression of the direct-detection
rate is tied to the pseudo-Goldstone nature of the $\chi$ field.

\begin{figure}[t]
  \centering
  \includegraphics[width=0.6\textwidth]{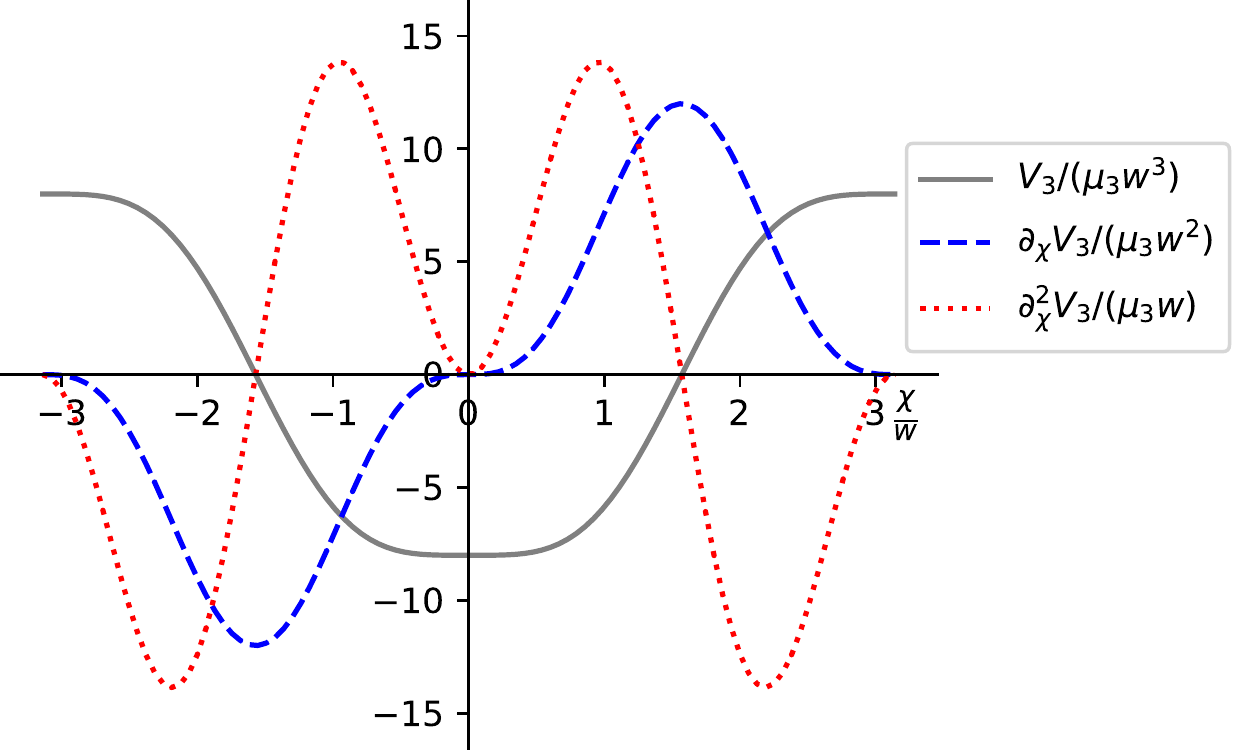}
  \caption{The potential, $V_3$, and its first two derivatives with
    respect to $\chi$ at $s=0$ for ${\mu_3^{\prime}=-9\mu_3}$.}
  \label{fig:ders}
\end{figure}

In the special case
$\mu_3^{\prime}=-9\mu_3$, when the tree-level direct-detection
cross section cancels, the contribution to $m_\chi$ from the cubic terms goes to zero as well, thereby
implying a more symmetric vacuum than just an accidental cancellation.
This can be understood by the form of the $\chi$-potential at the vacuum,  eq.~\eqref{eq:V3}. To illustrate this explicitly, we show the cubic part of the potential and its derivatives
for the special parameter domain $\mu_3^{\prime}=-9\mu_3$
in figure~\ref{fig:ders}.
As we noted above, the second derivative of $V_3$ with respect to $\chi$ vanishes when
evaluated in the vacuum.
However, notice that the cubic contributions along the $s$ direction
are not zero at $\chi=0$ even for $\mu_3^{\prime}=-9\mu_3$ implying
that the full Lagrangian does not have an enhanced symmetry in this limit.

In the general case of eq.~\eqref{DD coupling Niko}, setting the combinations of parameters which appear in the direct-detection cross section,
\be
9\mu_3 + \mu'_3, \qquad
4 \mu^3_1 + v^2 \mu_{HS}, \qquad
4 w \lambda'_{HS} + \mu_{HS},
\label{eq:cancellationregions}
\ee
to zero leads to suppression of the direct-detection rate.
It can be shown that these same combinations also appear in $m_\chi$.
To illustrate these conclusions, we show in the left panel of figure~\ref{fig:DD1L} the contours of the ratio of tree-level $\sigma_{\rm{SI}}$ to the XENON1T upper limit on the DM--nucleon cross section for the
parameter combination $9\mu_3 + \mu'_3$. Thus, contours with values of one or less are allowed. For this plot, we have chosen the typical values of $w=250$~GeV, $m_2=200$~GeV and $\sin\theta=0.1$, while all other symmetry-breaking terms (except for $\mu_{S}^{\prime 2}$) are set to zero.

\subsection{Direct-detection cross section at one loop}
\label{sec:DD1loop}

As shown in \cite{Azevedo:2018exj},
in the case of the simplest U(1)-invariant
model broken by only the $\mathbb{Z}_{2}$-symmetric mass term, the
non-vanishing corrections to the direct-detection cross section
in the $t\to 0$ limit arise at one-loop order.
When more general U(1)-breaking interactions are considered, they will
yield ${\cal O}(t^0)$ contributions to the direct-detection cross section
already at tree level. The allowed magnitude of these interactions
at tree level is expected to be roughly similar to the size of the loop
corrections due to the mass term~\cite{Alanne:2018zjm}. The loop corrections arising from
the symmetry-breaking interactions are then expected to be negligible.

However, as we have discussed above,
the contributions from the symmetry-breaking interactions are suppressed
at tree level in specific parts of the
parameter space. In such case the effect of loop corrections becomes
again relevant. Therefore, we will briefly discuss the one-loop contribution in the presence of these cubic terms extending the analysis
of ref.~\cite{Azevedo:2018exj}.

We will concentrate here on the case where the only non-zero cubics are $\mu_3$
and $\mu_3^{\prime}$. This choice is motivated by simplicity, but also because
we want to, in  particular, study if the suppression of the direct-detection cross section for the
specific choice $\mu_3^{\prime}=-9\mu_3$ is preserved at the loop level.

The spin-independent cross section with the one-loop corrections at $t\to 0$
limit can be written as
\begin{equation}
  \label{eq:DD1loop}
  \sigma_{\rm SI} = \frac{f_N^2m_N^2}{4\pi m_{\chi}^2}\left(\frac{m_\chi m_N}{m_\chi+m_N}\right)^2
    |\lambda_{\rm SI}^{\rm tree}+\lambda^{\rm 1L}_{\rm SI}|^2,
\end{equation}
where now $\lambda_{\rm SI}^{\rm tree}$ is given by eq.~\eqref{DD coupling Niko}
with $\mu_1=\mu_{HS}=0\,{\rm GeV}$, and $\lambda_{HS}^\prime=\lambda_S^\prime=0$.
In general, $\lambda_{\rm SI}^{\rm 1L}$ has a complicated analytic expression involving several loop functions. In the particular case $\delta\equiv 9\mu_3+\mu_{3}^\prime=0$ the expression simplifies significantly, and we show the result for illustration in appendix~\ref{app:DD1loop}.
In the numerical computation we keep the full expression including the deviation from the $\delta=0$ limit.

\begin{figure}
  \includegraphics[width=0.31\textwidth]{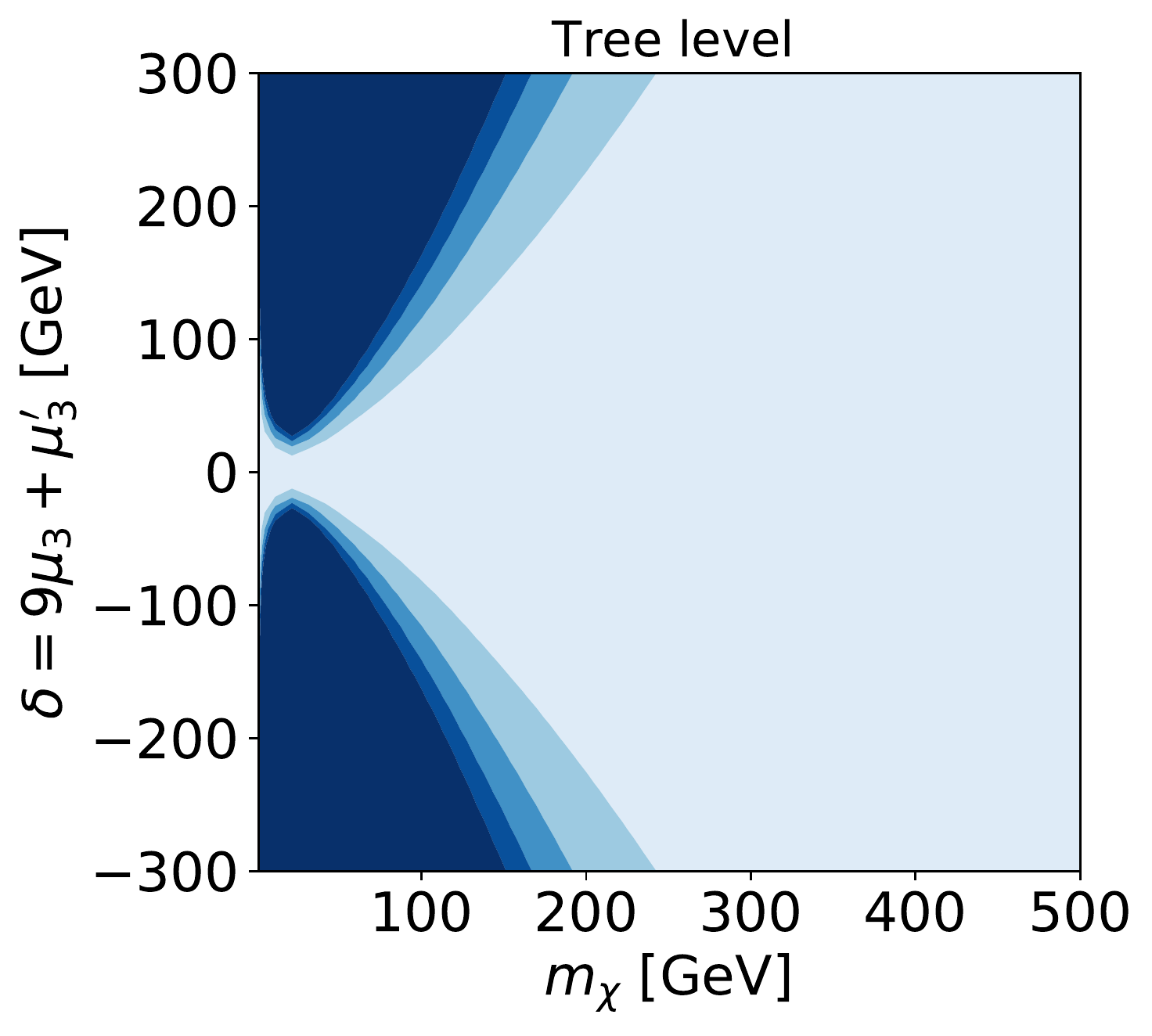}
  \includegraphics[width=0.31\textwidth]{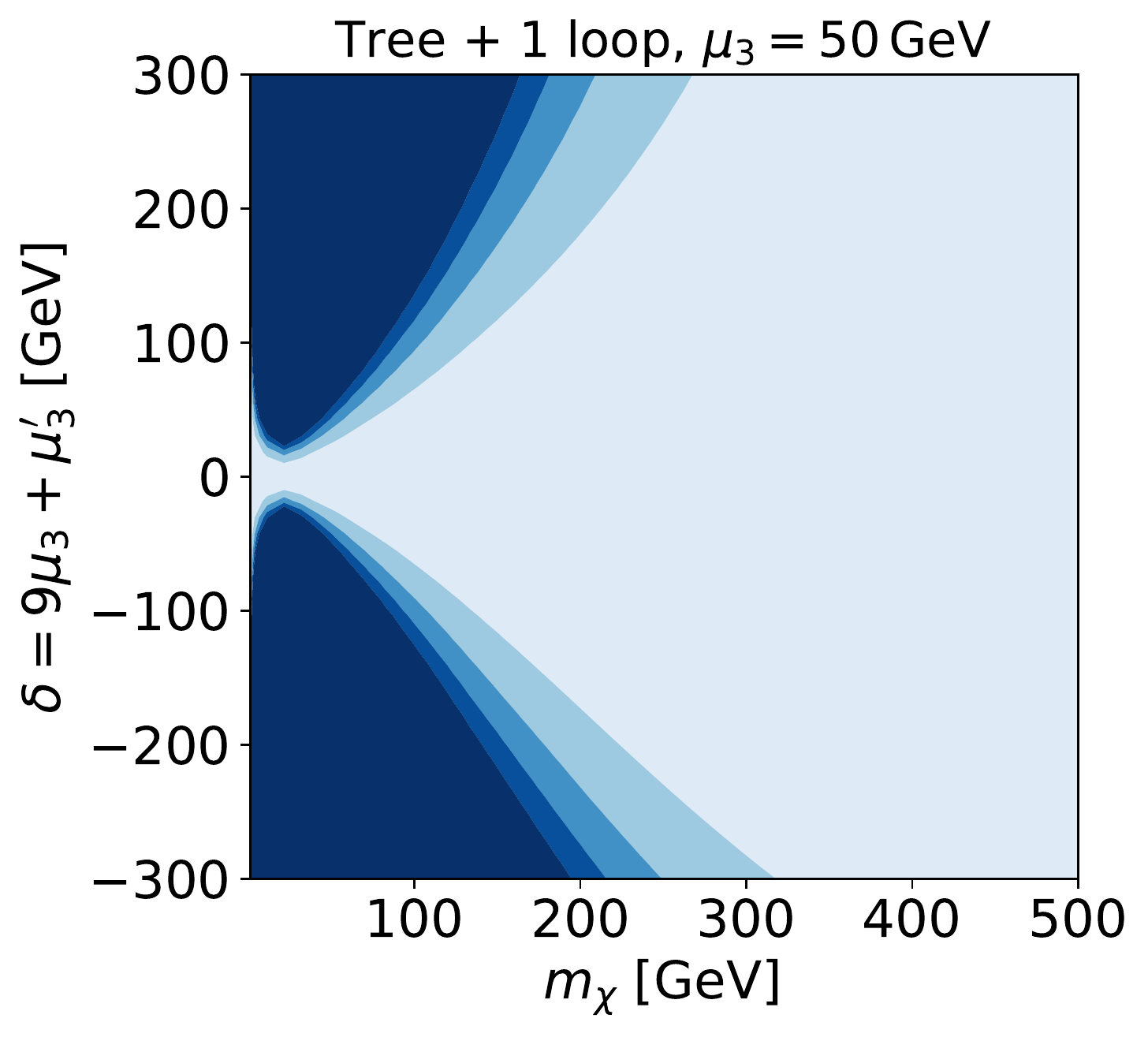}
  \includegraphics[width=0.365\textwidth]{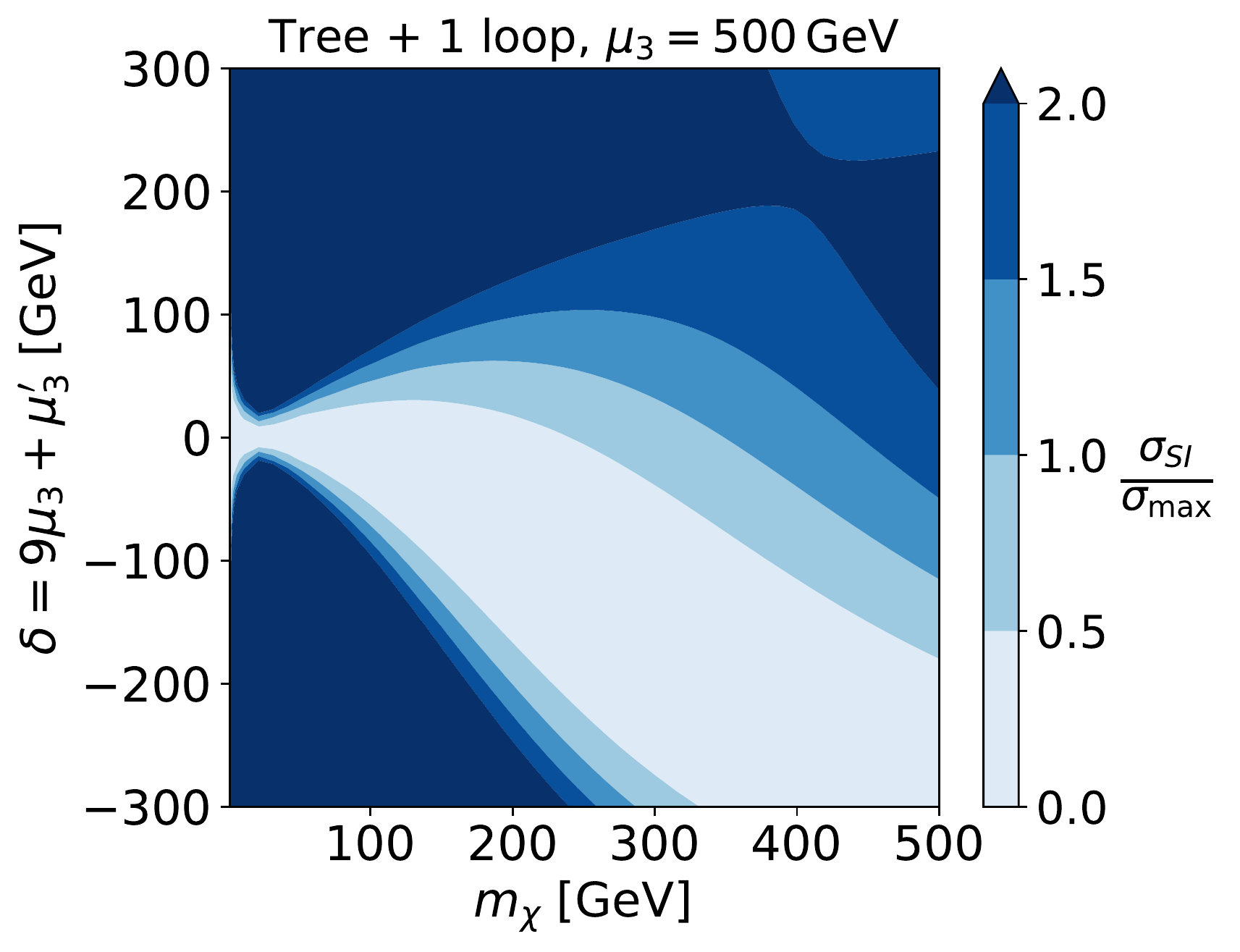}
  \caption{Cancellation regions for the DM-nucleon interaction cross section, regardless of the relic density contributions; contours show the value for $\sigma_{\rm SI}$ over the XENON1T limit cross section, $\sigma_{\max}$, meaning that the contours showing a value less than one pass the direct-detection bounds. Here, we have set $w=250$ GeV, $m_2=200$ GeV and $\sin\theta=0.1$. Note that all other symmetry-breaking terms not explicitly presented in the plot are set to zero: $\mu_1=\mu_{HS}= 0$~GeV and $\lambda'_S=\lambda'_{HS}=0$.}
  \label{fig:DD1L}
\end{figure}

While at tree level, the value of only the combination $\delta$ is relevant, at loop level also the individual values of the coupligns $\mu_3$ and $\mu_3^{\prime}$ become important, since the limit $\delta=0$ does not correspond to a symmetry at the
Lagrangian level. We illustrate this in figure~\ref{fig:DD1L}, where we plot the spin-independent cross section (again regardless of the relic-density contribution) at tree and one-loop levels as a function of $m_\chi$ and $\delta$ for two representative values of $\mu_3=50, 500$~GeV. For the numerical evaluation of the various loop functions,
we use the \texttt{pySecDec} toolbox~\cite{Borowka:2017idc,Borowka:2018goh} with \texttt{FORM} optimization~\cite{Vermaseren:2000nd,Kuipers:2013pba,Ruijl:2017dtg} and \texttt{CUBA} library for multi-dimensional integration~\cite{Hahn:2004fe,Hahn:2014fua}.

Figure~\ref{fig:DD1L} shows that the suppression of the direct-detection cross section persists also at one-loop level for moderate values of the couplings $\mu_3, \mu_3^{\prime}$, but at larger values the interference with the symmetry-breaking mass term increases the one-loop cross section significantly at large $m_\chi$. Note however, that near the $\delta=0$ region for $\delta<0$ and large $m_\chi$, the
tree-level and one-loop contributions interfere destructively, allowing for the region of
suppressed direct-detection cross section for larger DM masses, too.

We conclude that the situation is analogous to the simplest
$\mathbb{Z}_2$-symmetric model of pseudo-Goldstone DM:
starting from a model in the cancellation region of U(1)-breaking parameters
discussed in the previous section, we expect deviation from the symmetric
$t\to 0$ limit of the direct-detection cross section by contributions of the
order of the loop corrections discussed in this section.

Another feature arising from loop corrections is that the parameter combination
$\delta=9\mu_3+\mu_3^{\prime}$ may not stay zero under running of couplings.
The $\beta$-functions of the model are given in appendix~\ref{sec:rges}.
As a simple example, let us consider the case where all other symmetry-breaking
interactions are set to zero except $\mu_3$ and $\mu_3^\prime$. Then
we have
\begin{equation}
\begin{split}
 16 \pi^{2} \frac{d(9\mu_3 + \mu'_3)}{dt} &=
 12 \lambda_S (9 \mu_3 + \mu'_3) +8\lambda_S\mu'_3,
\end{split}
\end{equation}
which explicitly shows that the running of $\delta$ is not multiplicative.
Generally, the model should be viewed as a low-energy effective theory with
the coefficients of the symmetry-breaking operators taking non-zero values
constrained to be compatible with experiments and observations. Nevertheless,
the cancellation regions we have discussed here may be interesting towards
more complete model building, in particular with the relatively large regions of the parameter space near the $\delta=0$ limit with suppressed direct-detection cross section persisting even at one-loop level.

\subsection{Indirect detection}

Finally, to relate the present analysis to the results
of ref.~\cite{Alanne:2018zjm}, we discuss the implications of
indirect detection of DM for our model framework. The relevant indirect-detection constraints arise due to annihilation of DM in the dwarf galaxies that orbit our Milky Way. In these structures the DM is cold and therefore DM annihilations take place essentially at vanishing momentum: $s\to 4 m_\chi^2$. The annihilation cross section to all final states, $f\bar{f}$, $W^+W^-$, $ZZ$, $h_1 h_1$, $h_1 h_2$ and $h_2 h_2$ is non-vanishing in this limit (unless kinematically forbidden).  The annihilation cross sections for these processes are given in appendix~\ref{sec:x:sections}.

\begin{figure}
\begin{center}
\includegraphics[width=0.4\linewidth]{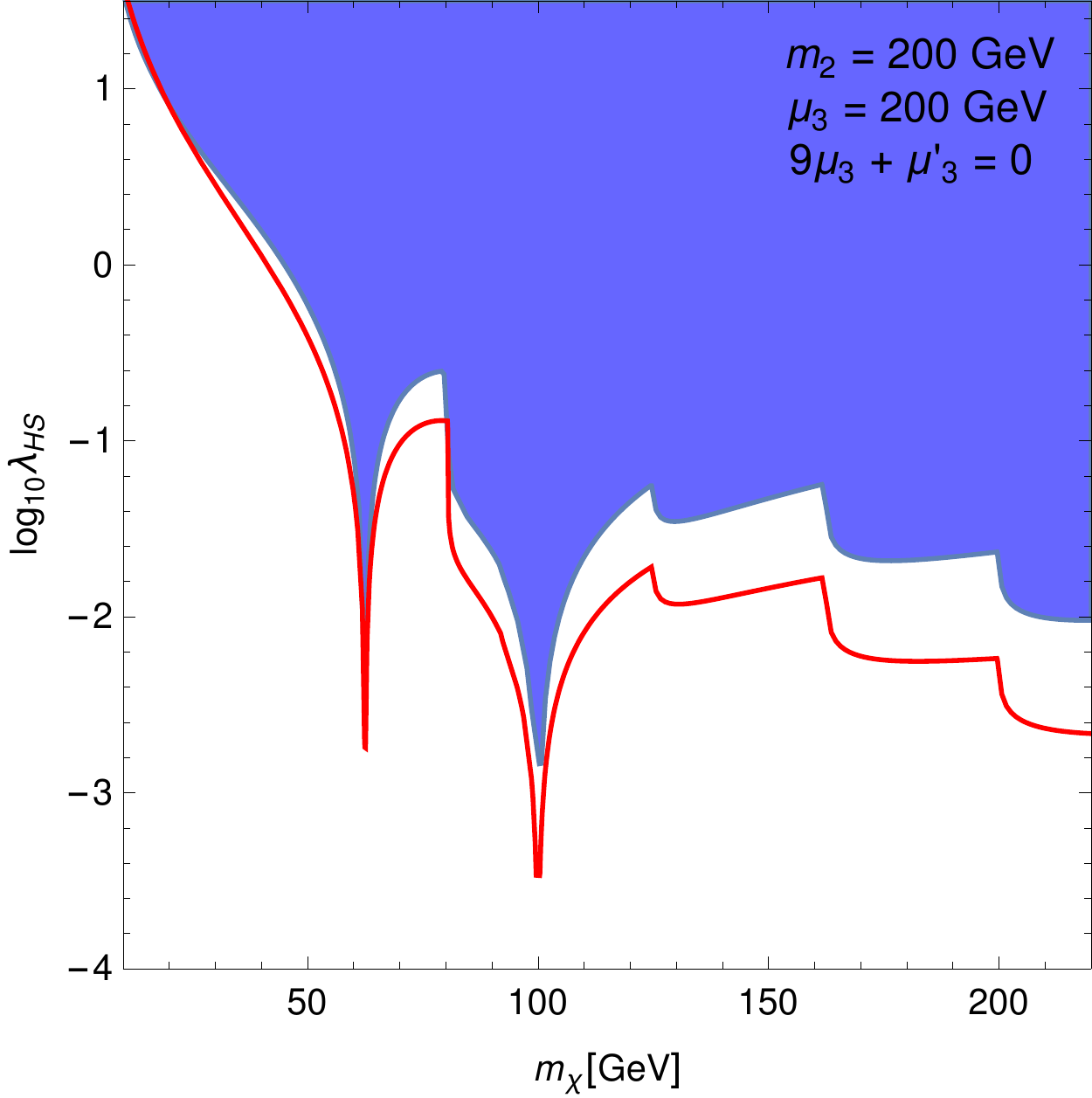}
\hspace{0.4cm}
\includegraphics[width=0.405\linewidth]{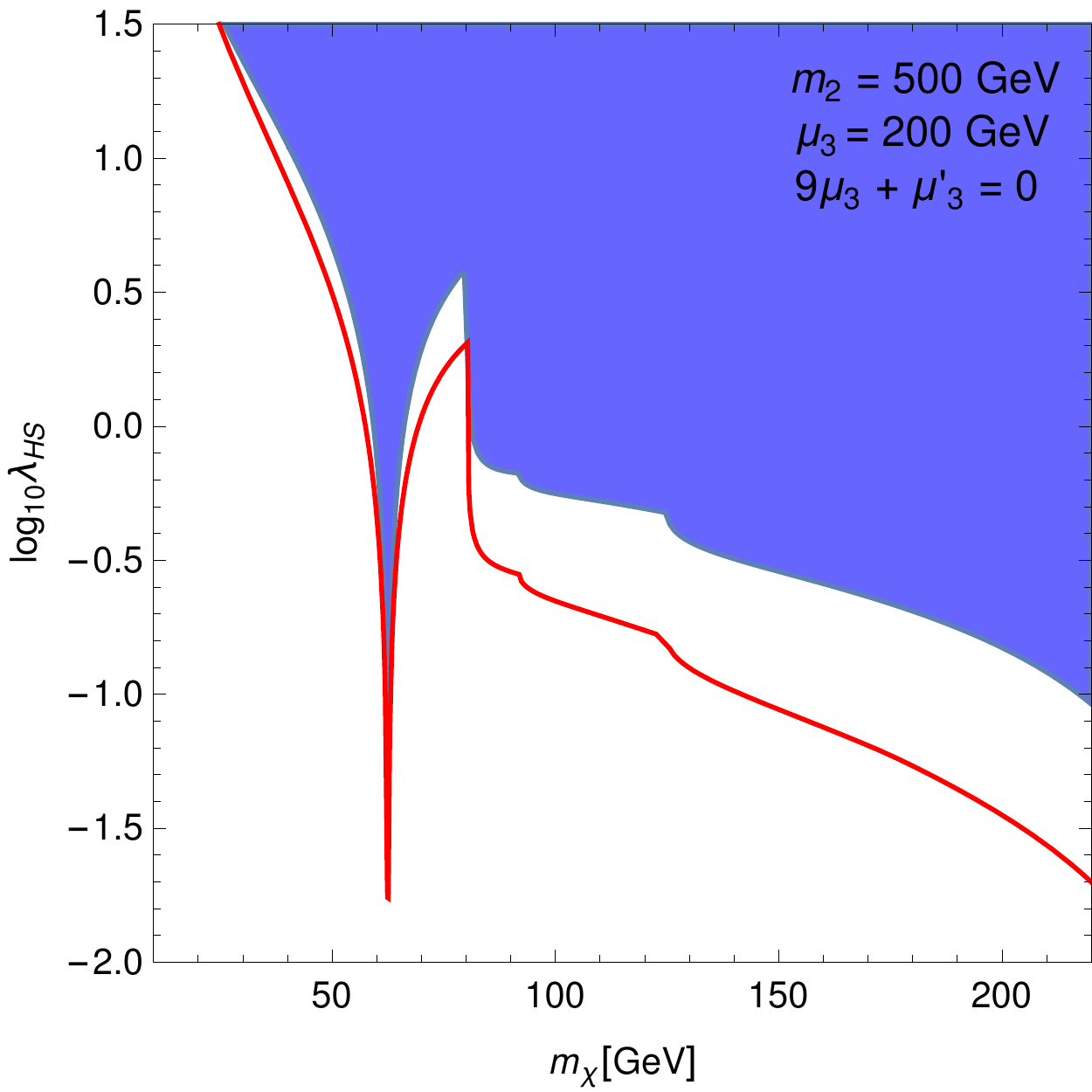}
\end{center}
\caption{The left panel presents the indirect-detection constraints for $m_{2}=200$ GeV and the right panel for $m_{2}=500$ GeV. In both panels  $\sin\theta=0.2$.
The red curve shows where the model reproduces the observed relic density,
$f_{\rm rel} = 1$.  The blue area is excluded by the latest Fermi-LAT data \cite{Boddy:2019qak}. Of the explicit breaking terms, only the mass term $\mu_S^{\prime 2}$ and the cubics $\mu_{3}$ and $\mu'_{3}$ satisfying
$\delta=9\mu_3+\mu_3^\prime=0$ are present.
}
\label{fig:indirect:detection:comparison}
\end{figure}

The constraints from the Fermi-LAT dwarf galaxy
observations, based on the recent analysis~\cite{Boddy:2019qak}, are presented in
figure~\ref{fig:indirect:detection:comparison}. The red curve shows where the model reproduces the observed relic density,
$f_{\rm rel} = 1$. The excluded regions, shown in blue, are
produced by comparing the annihilation cross section
$v_{\rm rel}\,\sigma_{\chi\chi\to \, b \bar{b}}$ to the reported  $b\bar{b}$ exclusion
limit for $m_\chi<m_{W^\pm}$.
In the region $m_\chi>m_{W^\pm}$, the dominant annihilation
channel is to $W^+W^-$ and in this region we compare the total DM annihilation cross
section\,---\,that is, the combined cross sections $\chi\chi\to W^+W^-, ZZ, h_1 h_1, h_1 h_2, h_2 h_2$\,---\,to the $W^+W^-$ bound.
The kinks in the plot occur at kinematic thresholds for each channel, that is at ${m_\chi=m_{W^\pm}, m_Z, m_{1}, (m_1+m_2)/2, m_2}$. We have set all the explicit
U(1)-breaking terms to zero in figure~\ref{fig:indirect:detection:comparison},
except for the mass term $\mu_S^{\prime 2}$ and the cubics $\mu_{3}$ and $\mu'_{3}$ satisfying $\delta=9\mu_3+\mu_3^\prime=0$.

The gamma ray flux originating from DM annihilations in a dwarf galaxy depends on the density profile of the DM halo. This effect is described via the so-called $J$-factor. In the present analysis we use the constraints on $\chi\chi\to b\bar{b}$  and $\chi\chi\to W^+W^-$ cross sections based on $J$-factors obtained in~\cite{Boddy:2019qak}. These updated bounds are weaker than the ones used in our previous work~\cite{Alanne:2018zjm},\footnote{In~\cite{Alanne:2018zjm}, the bounds on  $\chi\chi\to b\bar{b}$  and $\chi\chi\to W^+W^-$ cross sections were taken from  refs.~\cite{Fermi-LAT:2016uux,Clark:2017fum,Boddy:2018qur}.} and therefore we find the model less constrained by the indirect-detection data. We conclude that, taking into account the uncertainties in the determination of the $J$-factors, the model is not presently constrained by indirect detection in the $m_\chi>m_{1}/2$ region.

\section{Phase transitions and gravitational waves}
\label{sec:pt}

\subsection{Thermal potential}
So far we have seen how the pseudo-Goldstone DM model can
account for the observed relic abundance, and how it is constrained
by direct-detection experiments. Since the model consists of an
extended scalar sector, it is natural to explore the finite-temperature phase transitions in the early universe
and their phenomenological consequences, in particular for the
GW signals relevant for the LISA, BBO or DECIGO satellites.

It turns out that the phenomenologically viable scenario  is a two-step
transition starting from the high-temperature vacuum in the singlet direction
$(h,s)=(0,w_0)$. With the linear and the cubic terms present\,---\,in the absence of an
additional $\mathbb{Z}_2$ symmetry\,---\,the symmetry in the $s$ direction is not necessarily restored, and $w_0$ can be non-zero.
In the first step, as
temperature is lowered, another minimum forms in the singlet direction,
$(h,s)=(0,w_1)$, and the transition $(0,w_0)\to(0,w_1)$ occurs at the
critical temperature $T_{c}$. This transition is potentially of first order, and
can produce GW signals to be searched for in the future
space-based missions. The electroweak transition from $(0,w_1)$ to $(v_2,w_2)$
happens at a significantly lower temperature $T_{c}^{\rm EWPT}\ll T_{c}$, and finally evolves
to the zero-temperature global minimum $(v,w)$.
In the phenomenologically viable parameter space for DM, this second transition is
predominantly of second order, and does not produce detectable
GW signals. In the following, we will concentrate on the
former first-order transition.

Let us now turn to the quantitative analysis of this scenario.
The effective one-loop potential reads
\begin{equation}
\label{eq:Veff}
V_\mathrm{eff} = V_\mathrm{tree} + V_{\rm CW}^0 + V_{\rm 1L}^{T} + V_{\rm CT} \,,
\end{equation}
where the tree-level potential is given by eq.~\eqref{original_potential}.
The second term, $V_{\rm CW}^0$, is the $T=0$ Coleman--Weinberg potential in
the $\overline{\rm MS}$ scheme,
\begin{equation}
\label{eq:VCW}
V_{\rm CW}^0 = \frac{1}{64\pi^2}\sum_i (-1)^{F} g_i\, M^4_i(h, s) \left[\ln \frac{M^2_i(h,s)}{\mu_0^2}-C_i\right] ,
\end{equation}
where $g_i$ denotes the number of degrees of freedom,
$M_i(h, s)$ are the field-dependent masses, and $\mu_0$ is the
renormalisation scale (which we fix to be $\mu_0 = v$).
In this expression, $F=1$ for fermions and $0$ for bosons,
$C_i = 3/2$ for scalars, fermions and longitudinal polarizations of gauge bosons and $1/2$ for transverse polarizations of gauge bosons.

The one-loop finite-temperature corrections are given by
\begin{equation}
\label{eq:VT}
V_{\rm 1L}^T = \frac{T^4}{2\pi^2}\sum_i g_i\, J_{\pm}\left(\frac{M_i(h, s)}{T}\right), \qquad
J_{\pm}(x)=\pm\int_0^{\infty}\!\mathrm{d}y\ y^2\ln\left(1\mp \mathrm{e}^{-\sqrt{x^2+y^2}}\right) \,,
\end{equation}
where the upper signs correspond to bosons and lower signs to fermions.

The last term in eq.~\eqref{eq:Veff} contains the finite parts of the counter
terms that are fixed such that the scalar vevs and masses remain
at their tree-level values at the minimum, eq.~\eqref{eq:vacuum}:
\begin{equation}
  \label{eq:VCT}
  V_{\mathrm{CT}} = \delta\mu_H^2 |H^2|+ \delta\mu_S^2 |S|^2
  +\frac{1}{2}\delta\mu_S^{2\,\prime}(S^2+S^{*\,2}) +\delta\lambda_H|H|^4
  + \delta\lambda_S|S|^4+\delta\lambda_{HS} |S|^2 |H|^2\,,
\end{equation}
such that the following renormalization conditions are satisfied:
\begin{equation}
\label{eq:}
\left.\frac{\partial V_{\mathrm{CT}}}{\partial \varphi_i}\right|_{\mathrm{vac}}
= -\left.\frac{\partial V_{\rm CW}^0}{\partial \varphi_i}\right|_{\mathrm{vac}} \,, \quad
\left.\frac{\partial^2 V_{\mathrm{CT}}}{\partial \varphi_i\partial\varphi_j}\right|_{\mathrm{vac}}
= -\left.\frac{\partial^2 V_{\rm CW}^0}{\partial \varphi_i\partial\varphi_j}\right|_{\mathrm{vac}} \,,\quad \varphi=(h,s).
\end{equation}
Finally, we use the thermally improved finite-temperature potential, which is obtained by adding to the field-dependent masses in eqs.~\eqref{eq:VCW},~\eqref{eq:VT} the leading thermal corrections (see ref.~\cite{Kainulainen:2019kyp} for a recent discussion):
\begin{equation}
\label{eq:thermal_masses}
M_i^2(h, s) \rightarrow M_i^2(h, s) + c_i\,T^2,
\end{equation}
where the coefficients $c_i$ are given by
\begin{align}
c_h =& (9g^2 + 3g^{\prime\,2}+12y_t^2+24\lambda_H+4\lambda_{HS})/48,\nonumber\\
c_s =& (2\lambda_{HS}+2\lambda_{HS}^{\prime} + 4\lambda_S ) / 12,\\
c_{\chi}=& (2\lambda_{HS} - 2\lambda_{HS}^{\prime} + 4\lambda_S)/12.\nonumber
\end{align}

\subsection{Gravitational-wave signal and peak-integrated sensitivity curves}

During a first-order phase transition, stochastic GWs are produced via three independent
mechanisms: collisions of bubbles (b), sound waves in the plasma (s), and
turbulence in the plasma (t). The resulting GW spectra can be approximately
written in terms of a peak amplitude, $\Omega^{\rm peak}_i$, and a spectral shape, $\mathcal{S}_i$ which depends on the peak frequency, $f_i$,
\begin{equation}
\label{eq:GW_spectra_peak}
h^2 \Omega_i\left(f\right) = h^2 \Omega_i^\text{peak}\, \mathcal{S}_i(f, f_i) \,.
\end{equation}

The peak amplitudes and peak frequencies depend on the characteristics of the
phase transition which can be quantified
in terms of the nucleation temperature, $T_n$,
the amount of energy density released relative to the radiation energy density \cite{Ellis:2019oqb}
\begin{equation}
\alpha \equiv \frac{1}{\rho_\text{rad}} \left( \Delta V - \frac{T}{4} \Delta \frac{\mathrm{d}V}{\mathrm{d}T} \right)
\end{equation}
characterising the strength of the transition, and in terms of
\begin{equation}
  \beta/H_n \equiv T_n\frac{\D(S/T)}{\D T}
  \qquad \mbox{with} \quad H_n=H(T_n),
\end{equation}
which gives approximately the inverse duration of the transition \cite{Grojean:2006bp}. Here, $S$ is the Euclidean action of the bubble solution. The latent heat released during the
phase transition is converted with efficiency $\kappa_{\rm b}$  into the kinetic energy of the expanding bubbles, with $\kappa_{\rm s}$ into the sound waves, and with $\kappa_{\rm t}$ into the turbulent motion in the plasma.
There is still no consensus in the literature on how latent heat released into the
kinetic energy of the plasma, $1-\kappa_{\rm b}$, is subsequently transformed
into sound waves and turbulence \cite{Caprini:2009yp,Caprini:2015zlo, Alves:2018jsw,Axen:2018zvb,Ellis:2019oqb,Guo:2020grp}; here we choose the
commonly used estimate for the turbulence fraction $\kappa_{\rm t}=0.1$, and we fix the bubble wall velocity $\vw = 0.9$.

The peak amplitudes and peak
frequencies depend on parameters $\vw$ and $\kappa_i$ and on the
values of functions $\alpha$ and $\beta$
evaluated at the nucleation temperature;
explicit formulas for these in the runaway-bubble-in-plasma scenario
 are given in refs.~\cite{Caprini:2015zlo,Ellis:2019oqb} along with
the spectral shape functions, $\mathcal{S}_i$, in terms of the peak
frequencies.

The resulting GW signal needs to be compared with the noise spectrum of the experiment under consideration to obtain the signal-to-noise ratio (SNR)~\cite{Allen:1997ad,Maggiore:1999vm}
\begin{equation}\label{eq:SNR_def}
    \snr = \left[n_\text{det}\, \frac{t_\text{obs}}{s} \int_{f_\text{min}}^{f_\text{max}} \frac{\mathrm{d} f}{\Hz} \left(\frac{h^2\Omega_\text{signal}(f)}{h^2\Omega_\text{noise}(f)}\right)^2 \right]^{1/2} \,,
\end{equation}
where $n_{\rm det} = 1$ if the experiment consists of just one detector allowing
only auto-correlation measurement, while $n_{\rm det} = 2$
if it is possible to cross-correlate signals of a detector pair.
In the following, we will consider three satellite-borne GW interferometers:
LISA~\cite{Audley:2017drz,Baker:2019nia}, DECIGO~\cite{Seto:2001qf,Kawamura:2006up,Yagi:2011wg,Isoyama:2018rjb}, and BBO~\cite{Crowder:2005nr,Corbin:2005ny,Harry:2006fi} with the planned
configurations $n_{\rm det} = 1$ for LISA and $n_{\rm det} = 2$ for DECIGO
and BBO. The noise spectra of these experiments are discussed in detail in
ref.~\cite{Schmitz:2020syl}.

For the representation of the model parameter points in the GW signal region
and the experimental reach of the above interferometers, we adopt the
approach of peak-integrated sensitivity (PIS) curves put forward recently in
refs.~\cite{Alanne:2019bsm,Schmitz:2020syl}. The advantage of this approach with
respect to
the conventional power-law-integrated sensitivity curves~\cite{Thrane:2013oya}
is that it allows to represent each parameter point as a single point in the
GW signal region and is thus well-suited for the purpose of a general scan
of the parameter space to be discussed in the following section.

The key observation is that if the shape of the expected signal is known, as
is the case of first-order phase transitions, the integration over the
spectral shape can be carried out leaving SNR uniquely determined by the peak
energy densities and peak frequencies that depend on the model-specific
phase-transition quantities and no longer on the GW frequency. More specifically,
the SNR in eq.~\eqref{eq:SNR_def} can be rewritten as
\begin{equation}\label{eq:PISC_SNR}
    \begin{aligned}
        \frac{\snr^2}{t_\text{obs}/\mathrm{yr}} =& \left(\frac{h^2\Omega^\text{peak}_\text{b}}{h^2\Omega_\text{PIS}^\text{b}}\right)^2    +
        \left(\frac{h^2\Omega^\text{peak}_\text{s}}{h^2\Omega_\text{PIS}^\text{s}}\right)^2    +
        \left(\frac{h^2\Omega^\text{peak}_\text{t}}{h^2\Omega_\text{PIS}^\text{t}}\right)^2    \\
    & + \Bigg(\frac{h^2\Omega^\text{peak}_\text{b/s}}{h^2\Omega_\text{PIS}^\text{b/s}}\Bigg)^2 +
        \Bigg(\frac{h^2\Omega^\text{peak}_\text{s/t}}{h^2\Omega_\text{PIS}^\text{s/t}}\Bigg)^2 +
        \Bigg(\frac{h^2\Omega^\text{peak}_\text{b/t}}{h^2\Omega_\text{PIS}^\text{b/t}}\Bigg)^2 \,,
    \end{aligned}
\end{equation}
where the integration over the frequency range has already been carried out
implicitly:
\begin{equation} \label{eq:def_PISC}
    h^2 \Omega_\text{PIS}^{i/j} \equiv \left[(2 - \delta_{ij}) \, n_\text{det} \, 1\, \mathrm{yr} \int_{f_\text{min}}^{f_\text{max}} \mathrm{d} f\:\frac{{\mathcal{S}}_i(f) \, {\mathcal{S}}_j(f)}{\left( h^2 \Omega_\text{noise}(f) \right)^2}\right]^{-1/2} \,,
\end{equation}
where $i,j \in \left\{\textrm{b},\textrm{s},\textrm{t}\right\}$ and
the mixed peak amplitudes are defined as geometric means,
\begin{equation}
h^2\Omega^\text{peak}_{i/j} = \left(h^2\Omega^\text{peak}_i\, h^2\Omega^\text{peak}_j \right)^{1/2} \,.
\end{equation}
For details of the approach, see refs.~\cite{Alanne:2019bsm,Schmitz:2020syl}.


\subsection{Scan of the parameter space and results}

We scan the parameter space with non-zero soft-breaking mass and
cubic interactions terms $\mu_3$ and $\mu_3^{\prime}$. We set the
quartic couplings $\lambda'_{S}$ and $\lambda'_{HS}$ to zero since they need
to be very small in any case to satisfy the XENON1T direct-detection bounds. We
also set $\mu_{HS}$ to zero as it does not contribute significantly to
producing a first-order phase transition in the
singlet direction in which we are interested.
We also set the linear $\mu_{1}^{3}$ term to zero
to reduce unnecessary degeneracy, since it can be eliminated in favour of the cubic couplings, cf. appendix~\ref{sec:shift:param}.
The parameter ranges we consider are $m_{\chi} \in [30, 500]$ GeV,
$\abs{\sin \theta} \in [0.01, 0.2]$, $m_{2} \in [10, 1500]$ GeV,
$\abs{\mu_{3}}, \abs{\mu'_{3}} \in [0, 250]$ GeV and
$\mum =\lambda'_{S}=\lambda'_{HS}=0$. To keep the scan denser, we employ a conservative perturbativity bound of $\pi/2$ on the absolute values of the quartic couplings.

To implement the constraints from the relic density, we use the \texttt{micrOMEGAs}
code~\cite{Belanger:2018mqt} with model files generated
by the \texttt{FeynRules}
package~\cite{Christensen:2008py,Christensen:2009jx,Alloul:2013bka}.
To search for the cosmological phase transitions and compute the nucleation
temperature, the tunnelling action between the vacua, as well as the
phase-transition quantities $\alpha$ and $\beta$, we employ the
\verb|CosmoTransitions|~\cite{Wainwright:2011kj} code. These results were also
checked by our own \texttt{Mathematica} code and using the \texttt{FindBounce} package \cite{Guada:2020xnz}.

In addition, we perform a separate scan to investigate the effect of the
linear term $\mu_1^3$.
In this case, we set $\mu_{3} = \mu'_{3} = \mu_{HS} = 0$, so they
are generated solely from elimination of the tadpoles.
This elimination is achieved by shifting the field
$s \to s + \sigma$ and then demanding that the linear
term of the resulting potential vanishes.
The effect of the shift on scalar
potential parameters is given in appendix~\ref{sec:shift:param}.
We parametrise this scan in terms of $\sigma$, not $\mu_{1}^{3}$ to avoid solving a cubic equation. In this scan we
consider parameter ranges $\abs{\sigma} \in [0.1,200]$~GeV, $\abs{\lambda'_{S}} \in [0, 0.001]$, $\abs{\lambda'_{HS}} \in [0, 0.01]$.
We consider the singlet vev in the range $|w| \in [1, 1.5 \times 10^{5}]~\text{GeV}$ and use it to fit the relic density to the observed value of $\Omega_{c} h^{2}$ within three standard deviations \cite{Aghanim:2018eyx}.
However, this linear term scan does not yield any points with strong first-order phase transition, nor is it particularly distinguishable in the direct-detection plots. For that reason we do not further discuss it separately.

\begin{figure}[t]
\begin{center}
\includegraphics[scale=0.5]{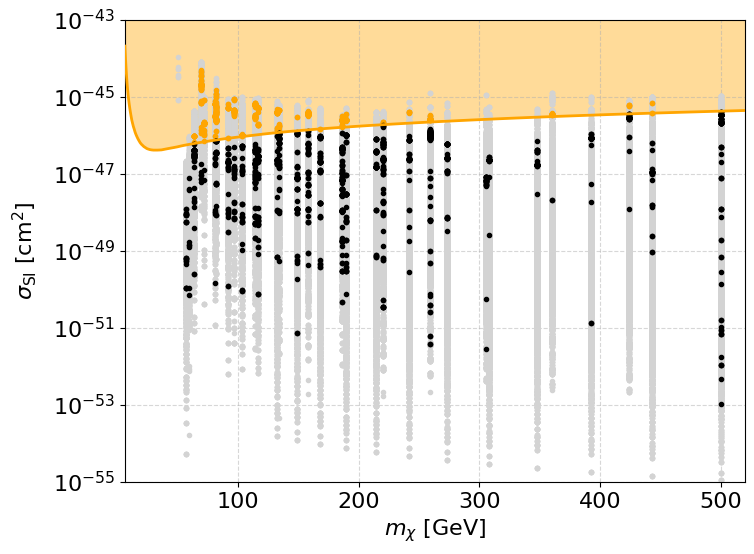}
\caption{The direct-detection cross section vs. DM mass where the orange shaded region is excluded by the XENON1T experiment. The black and orange points produce a first-order phase transition while the gray points fail to do so. The black
points are allowed by the XENON1T bound while the orange points are excluded.
}
\label{fig:direct:det}
\end{center}
\end{figure}

The results of the scan in the plane of direct-detection cross section vs. DM mass are shown in figure~\ref{fig:direct:det}. The points shown satisfy
all theoretical and experimental constraints discussed in
section~\ref{sec:model} except for the direct-detection bounds by XENON1T which rule out the orange shaded region.
The black and orange points produce a first-order phase transition while the gray points fail to do so. The black
points are allowed by the XENON1T bound while the orange points are excluded.
As is evident from the figure, there are ample regions of the parameter space
which provide a viable DM candidate and lead to a first-order
phase transition. For these sets of parameters, we then determine the
magnitude of the GW signal.

\begin{figure}[t!]
\centering
\includegraphics[width=0.49\textwidth]{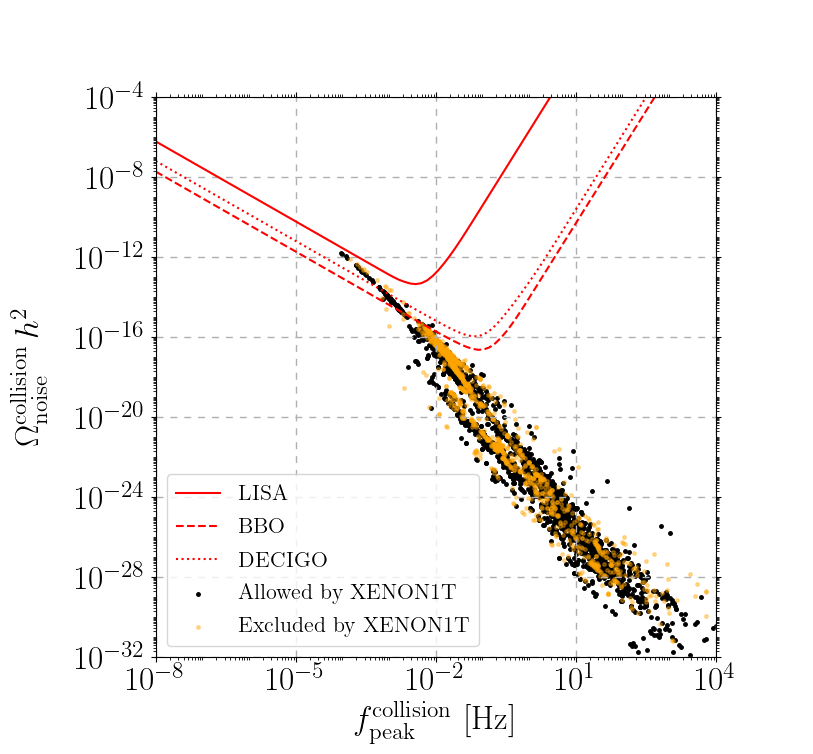}
\includegraphics[width=0.49\textwidth]{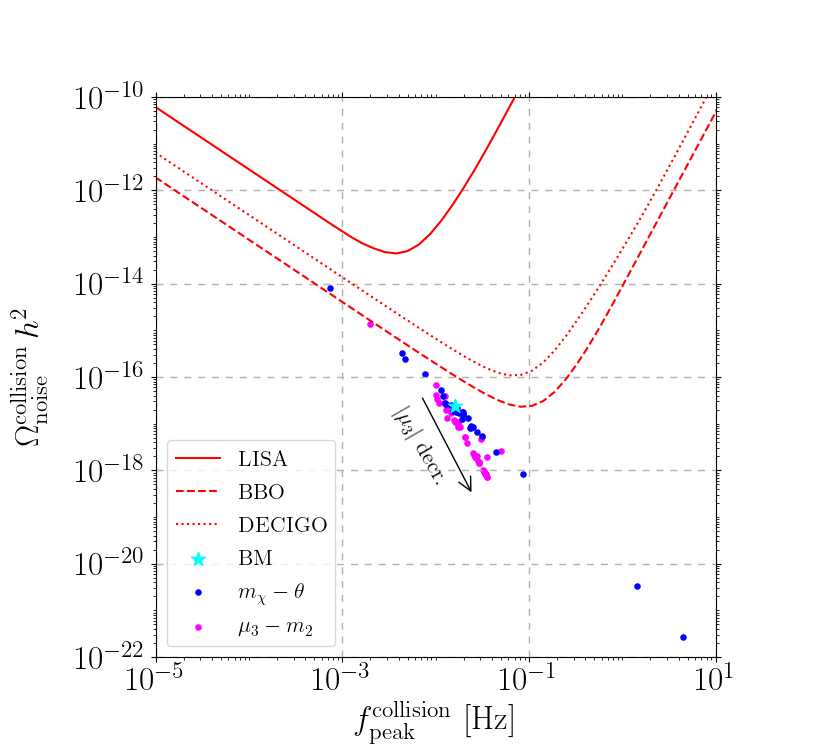}
 \caption{{\bf Left panel:} Points allowed by the XENON1T experiment in black and points ruled out by it in orange. {\bf Right panel:} Refined scan around a benchmark (BM) point along the $f_{\rm rel}=1$ hypersurface.
The blue points are obtained by varying $m_\chi$ and $\theta$
while the magenta points are the result of varying $\mu_3$ and $m_2$ values.
}
  \label{fig:GW_coll}
\end{figure}

For the stochastic GW background signal, we
recast the parameter points into the peak frequency--peak
energy density plane, fixing $\kappa_{\rm t}=0.1$. For the PIS curves, we assume the observational time
$t_{\rm obs} = 4~{\rm yr}$ and threshold SNR
$\rho_{\rm thr}=10$~\cite{Caprini:2015zlo,Audley:2017drz} for all
the experiments. The most sensitive channel turns out to be bubble collisions for which we show the parameter points in the GW signal region
in the left panel of figure~\ref{fig:GW_coll}. The black (orange) points
are allowed (excluded) by XENON1T constraints.

We observe that obtaining parameter sets which lead to strong enough
GW signal observable by future experiments becomes difficult
in the generic parameter scan described
in the beginning of this section.
This is typical for a multi-dimensional parameter space constrained by multiple observables: the
viable parameter space becomes more concentrated on lower-dimensional
hypersurfaces and refined parameter scanning is needed.

\begin{figure}
\centering
\includegraphics[width=0.5\textwidth]{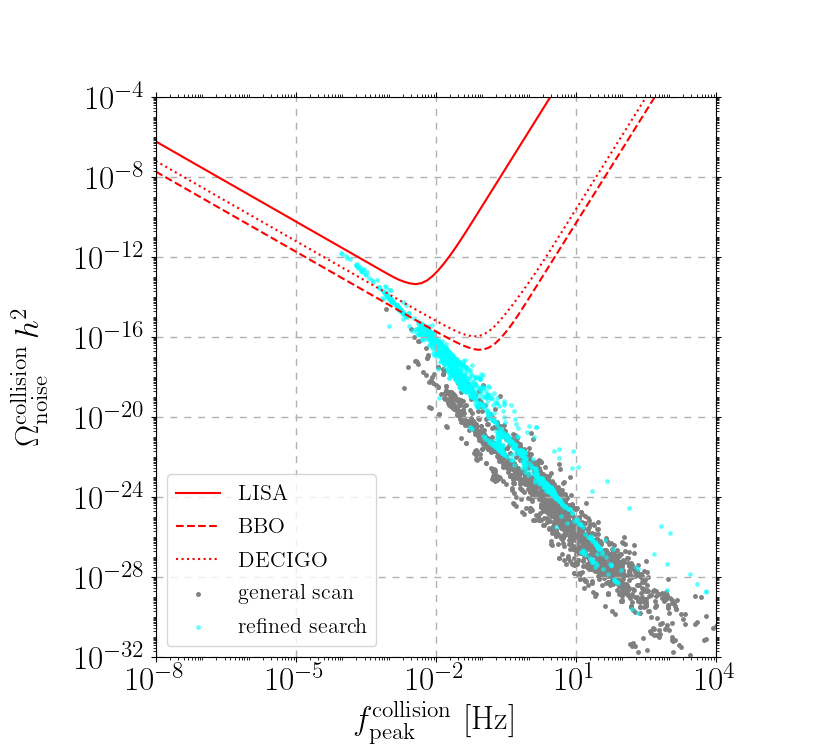}
\caption{The points from the general scan are shown in gray while
the points generated by refined scanning are shown in cyan.}
\label{fig:GW_coll-BM}
\end{figure}

Here, to explore the parameter space for the first-order phase transition more
closely, we choose a few viable benchmark points from the general scan and
search for more points in their vicinity along $f_{\rm rel}=1$
hypersurfaces varying only two parameters at a time. This search strategy is
illustrated in the right panel of figure~\ref{fig:GW_coll},
where we show the variation around a benchmark point
\[
(m_2,m_\chi, \theta,w, \mu_3, \mu_3^{\prime})=(400\,{\rm GeV},360\,{\rm GeV},0.0316, 1320\,{\rm GeV},-200\,{\rm GeV},-3.76\,{\rm GeV})
\]
by varying
either $(\mu_3,m_2)$ or $(m_{\chi},\theta)$. With this refined scan
we are able to obtain parameter sets which correspond to models with
sufficiently strong first-order transition to be visible in future
searches for GWs. To better show the effect of the refined
search over the full set of generated points, in figure~\ref{fig:GW_coll-BM}
we show the points from the general scan discussed earlier in gray and
the points generated by refined scanning in cyan.

\begin{figure}[t]
\includegraphics[width=0.5\textwidth]{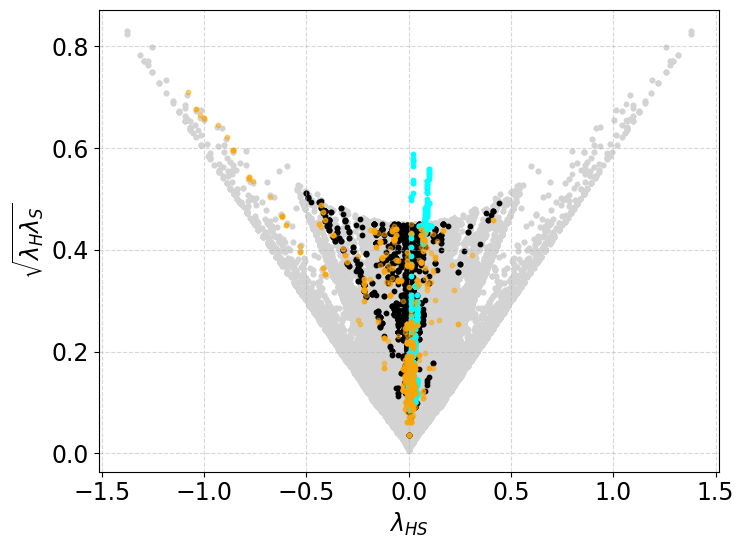}\
\includegraphics[width=0.48\textwidth]{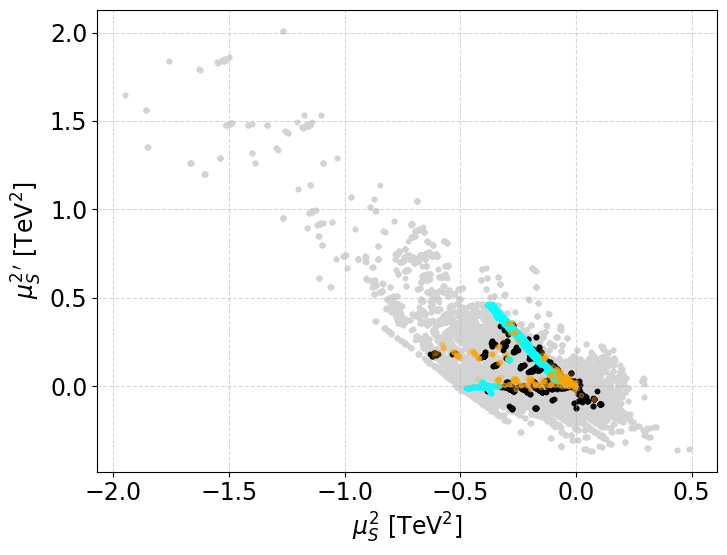}
\caption{The gray points in both panels show all
scanned points while the orange (black) points show the points
leading to a first-order phase transition but are excluded (allowed) by XENON1T constraints on the direct-detection cross section. The cyan points are the ones generated by the refined scan.}
\label{fig:scatter_pars_tweaked}
\end{figure}

To complete this discussion, we show how the
scanned points are distributed by looking at various projections
in the parameter space of the model.
In figure~\ref{fig:scatter_pars_tweaked}
the gray points in both panels show all
scanned points, while the orange (black) points show the points
leading to a first-order phase transition but are excluded (allowed) by XENON1T constraints on the direct-detection cross section. The cyan points are the ones generated by the refined scan.

In the left panel we show the dependence of the scanned points on
the portal coupling $\lambda_{HS}$ and on the combination
$\sqrt{\lambda_H\lambda_S}$.
The linear envelopes arise from the stability of the potential condition,
while the curve bounding the points from above is due to the
upper limit $\lambda_S=\pi/2$ set by hand to guarantee a conservative bound on perturbativity and unitarity.
As expected, the refined points cover very
specific regions in the parameter space. Note that in the left panel
some of the newly generated points go above the enveloping curve of the
general scan. This is merely due to releasing the constraint
$\lambda_S\le \pi/2$ slightly but without endangering unitarity.
The right panel shows the scanned points with respect to $\mu_S^2$
and $\mu_S^{\prime 2}$.

\begin{figure}
\includegraphics[width=0.32\textwidth]{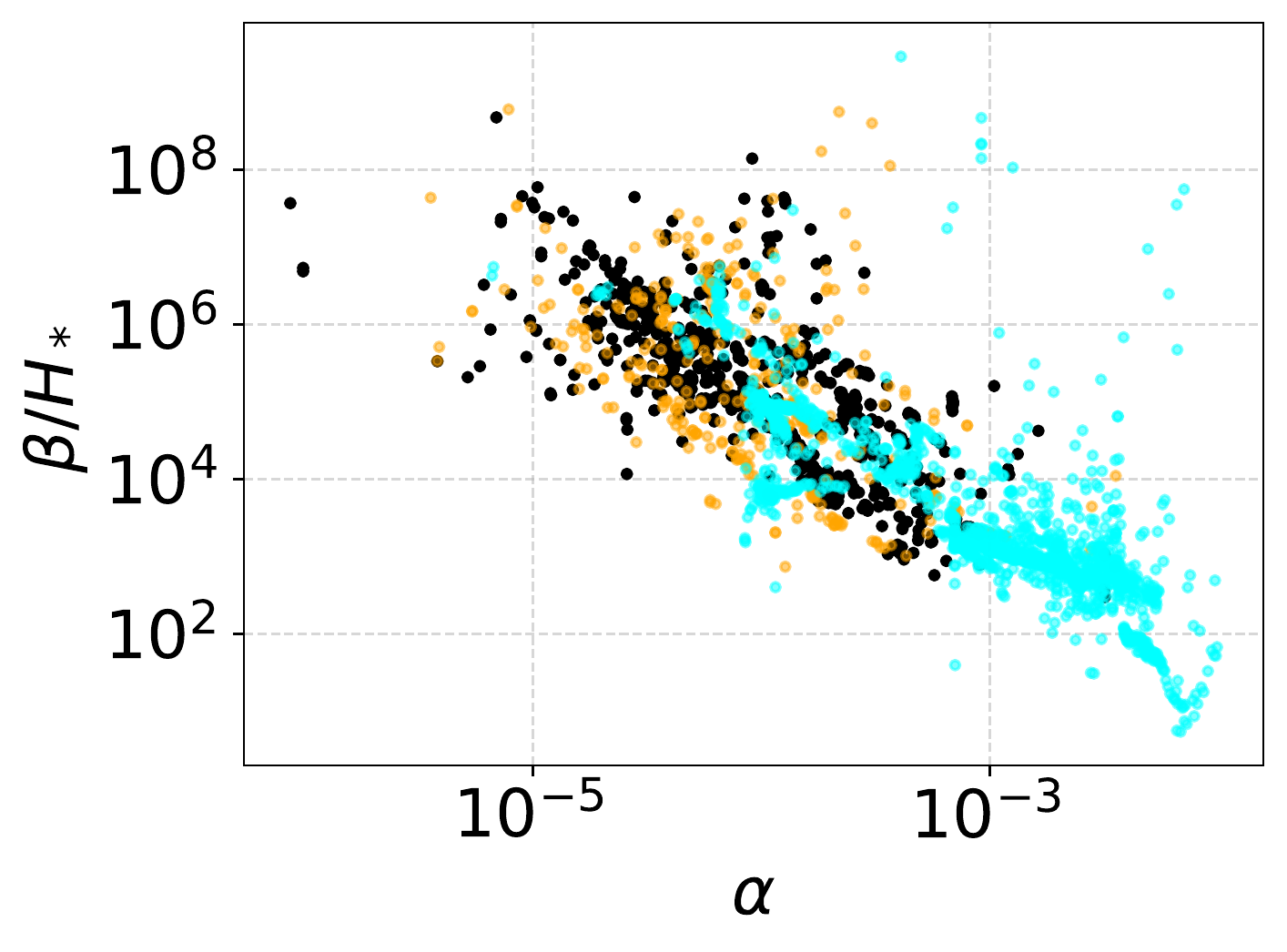}\
\includegraphics[width=0.32\textwidth]{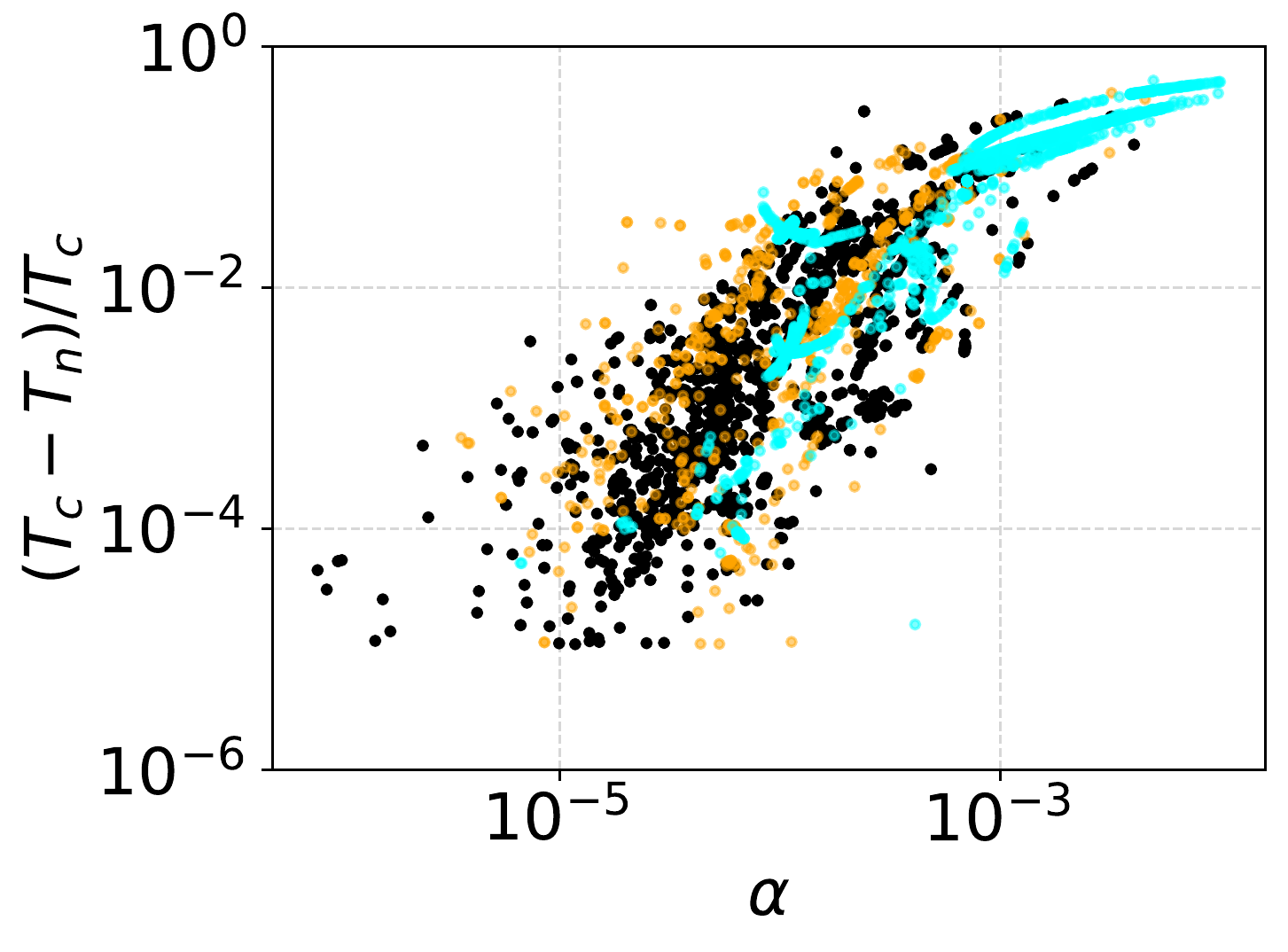}\
\includegraphics[width=0.32\textwidth]{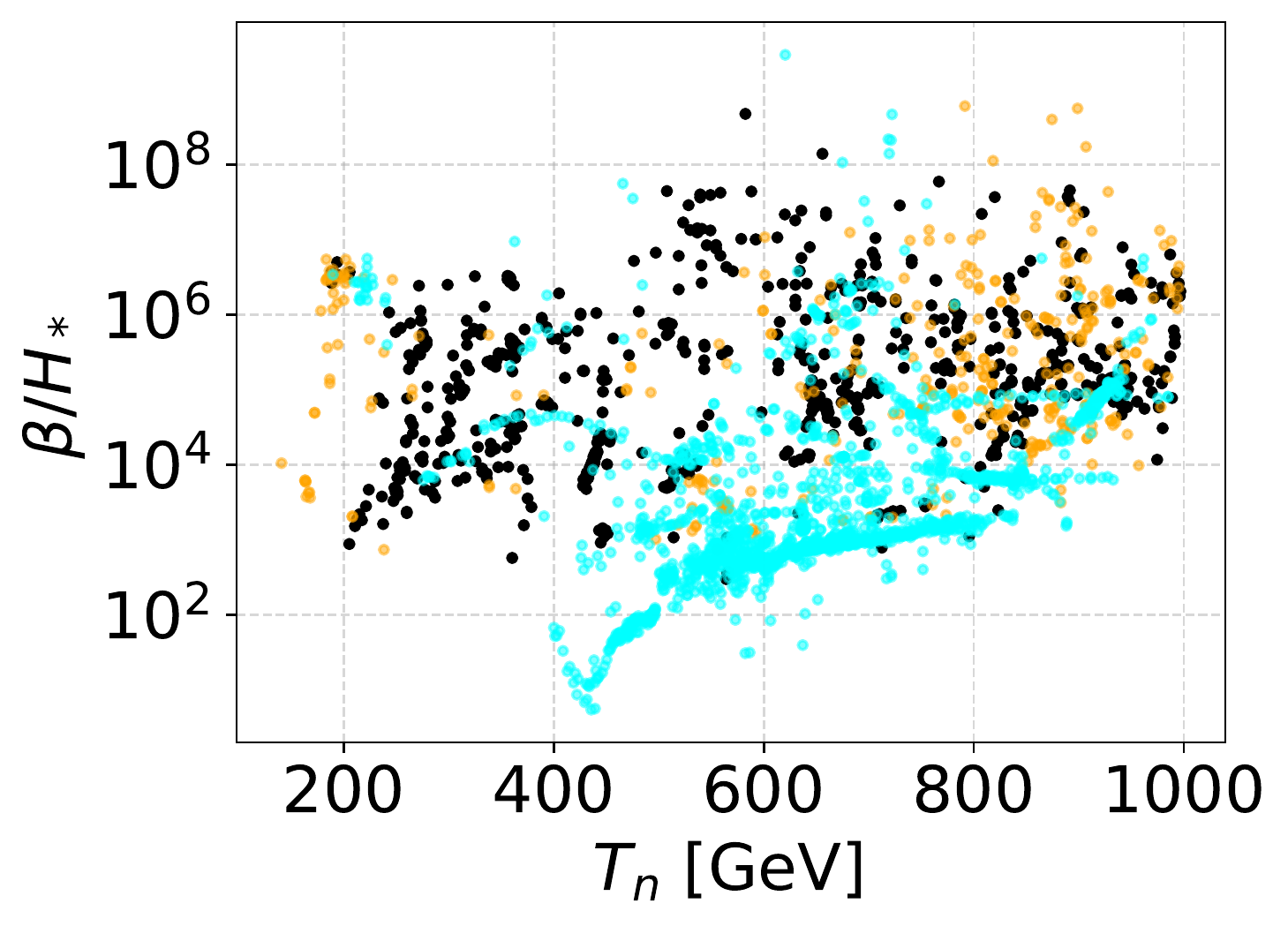}
  \caption{All points leading to a first-order phase transition including the points from the refined scan shown in cyan. The orange (black) points are excluded
(allowed) by the XENON1T experiment. Left panel: $\beta/H_n$ vs. $\alpha$
  at nucleation temperature. Middle panel: $(T_{c} - T_{n})/T_{c}$ vs. $\alpha$. Right panel: $\beta/H_n$ vs $T_{n}$.}
  \label{fig:scatter_PT_tweaked}
\end{figure}

In figure~\ref{fig:scatter_PT_tweaked} we show only the points which lead
to a first-order phase transition. The orange (black) points are excluded
(allowed) by the XENON1T experiment, while all other theoretical and experimental
constraints discussed in section~\ref{sec:model} are satisfied.
The cyan points again correspond to the refined scan.
The points are projected into the plane
of the quantities relevant for the GW signal. From the plots we see that
both $\beta$ and $1-T_n/T_c$ quantifying the amount of supercooling are correlated with $\alpha$, but the value
of $\beta$ and the value of the nucleation temperature are uncorrelated.
The refined points follow the same correlation pattern,
but the new points are more concentrated towards the region
of large~$\alpha$.

\begin{figure}
\includegraphics[width=0.5\textwidth]{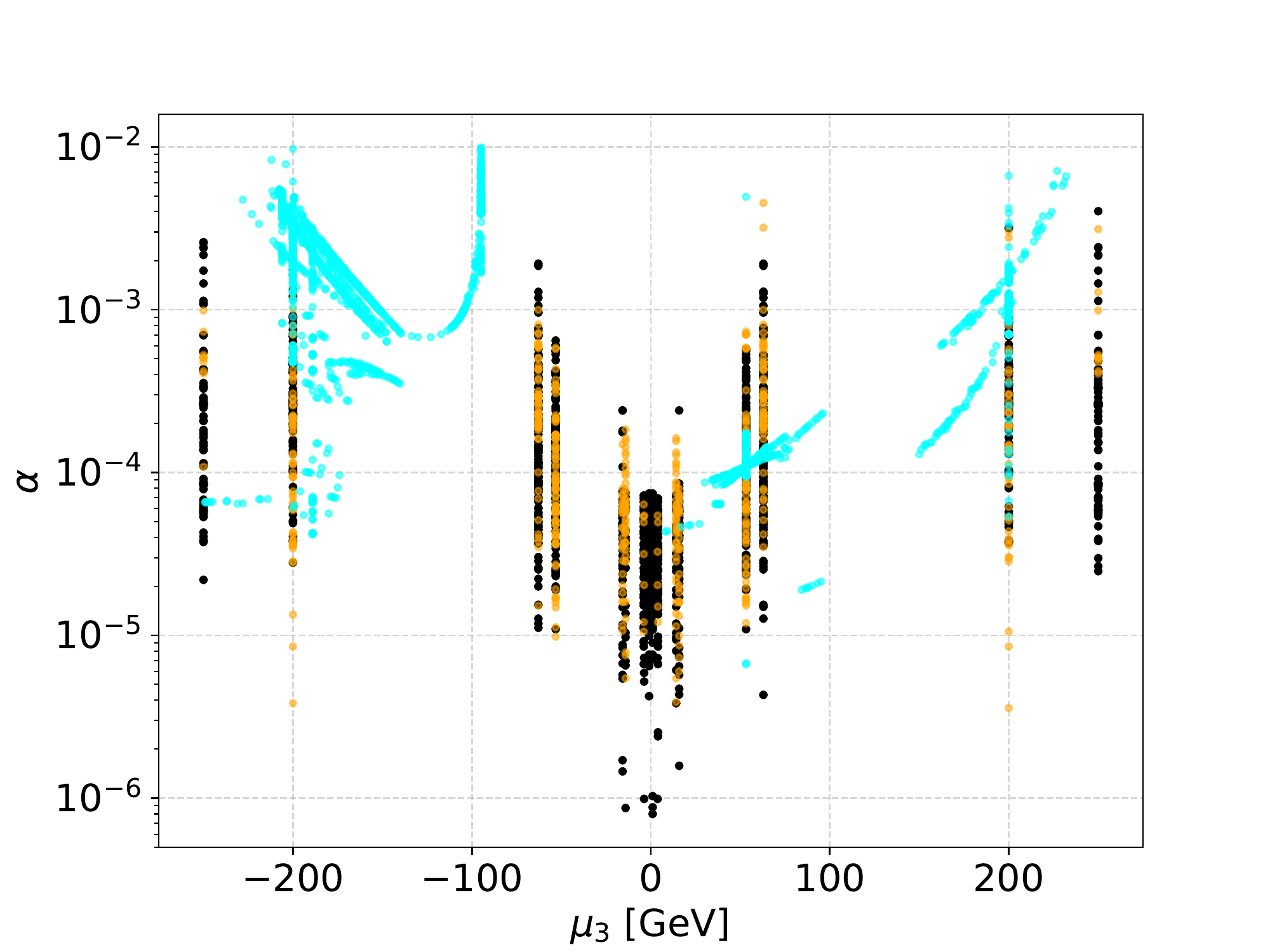}\
\includegraphics[width=0.5\textwidth]{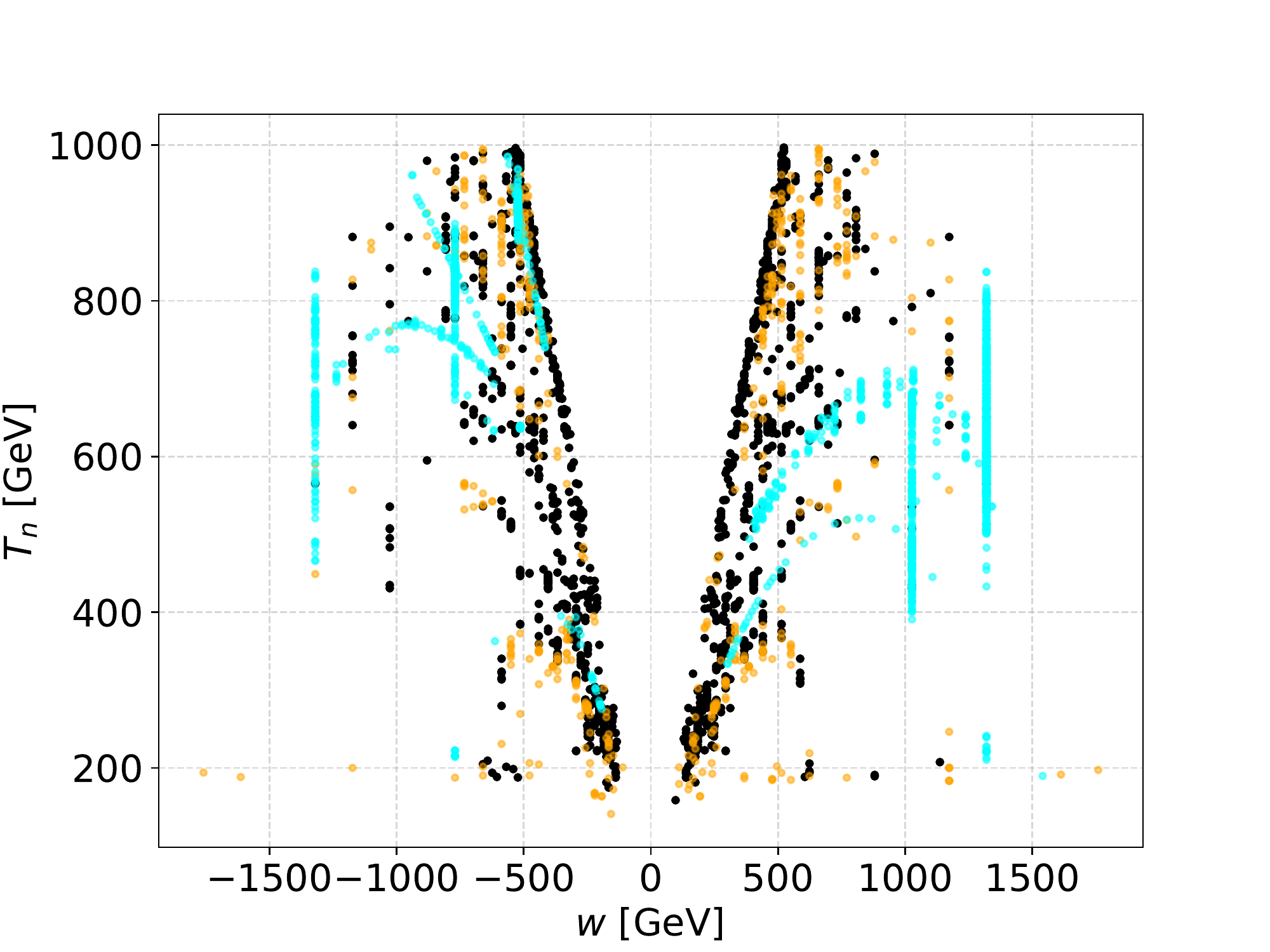}
  \caption{Dependence of phase transition parameters on potential parameters. The points from the refined scan are shown in cyan. The orange points are excluded and the black points are allowed by the XENON1T constraints. All other constraints are satisfied.}
\label{fig:scatter_pars_PT_tweaked}
\end{figure}

In figure~\ref{fig:scatter_pars_PT_tweaked} the same points as in
figure~\ref{fig:scatter_PT_tweaked} are shown, but
illustrating the dependence of GW signal on the parameters of the potential. In the left panel we
see that the non-zero value of $\mu_3$ allows for larger values of $\alpha$.
This is expected since a non-zero $\mu_3$ contributes to a stronger phase
transition. Similarly in the right panel we see that larger nucleation
temperature requires a larger absolute value for the vev of the singlet field, $|w|$.

This concludes our analysis: we have established the parameter space of the model
which provides for the observed DM abundance and
is compatible with all present constraints from collider searches and
direct and indirect DM detection. Moreover, this same parameter
space allows for a first-order phase transition in the early universe
whose resulting GW signal may be discovered in future
observations.

\section{Conclusions}
\label{sec:conclusions}

In this paper, we have considered the most general model of pseudo-Goldstone DM
that arises from the complex-singlet extension of the SM. Since the global U(1)
symmetry is completely broken, the only remaining discrete symmetry is
$S \to S^{*}$ which stabilises the imaginary part of the singlet as a DM candidate.

In the $\mathbb{Z}_{2}$-symmetric case, in which the only U(1)-breaking
term present is the mass term $\mu_{S}^{\prime 2}$, the tree-level direct-detection
cross section vanishes in the $t \to 0$ limit.
In the general case, we show that
all other U(1)-breaking terms give a non-zero contribution
to the direct-detection cross section in the $t\rightarrow 0$ limit
significantly increasing the interaction rates relevant for direct-detection
experiments and lifting the protection due to the pseudo-Goldstone
properties of the DM candidate.

However, we discovered that the symmetry-breaking parameters appear in
certain combinations, given explicitly in eq.~\eqref{eq:cancellationregions},
both in the pseudo-Goldstone mass and the tree-level direct-detection cross section.
Setting such
combinations to zero leads to cancellations which restore the pseudo-Goldstone
properties of the DM candidate and suppress its direct-detection
cross section in the $t\rightarrow 0$ limit.
Although the running of the couplings upsets this cancellation, the loop-level contributions
can be mild and keep the direct-detection cross section moderately suppressed and still
interesting from a phenomenological point of view as shown in figure~\ref{fig:DD1L}.
We also considered constraints from indirect detection, but
found that they do not presently constrain the parameter space significantly;
see figure~\ref{fig:indirect:detection:comparison}.

We also calculated the finite-temperature effective potential of the model
and study the implications for phase transitions in the early universe.
We considered a scenario where there is first a first-order transition
in the singlet direction $(w,v)=(w_0,0)\to(w_1,0)$ with zero Higgs vev $v$,
from which a second-order
transition brings the fields to the electroweak minimum where both
the singlet and the Higgs have non-zero vevs. The barrier in the first
transition is generated by the singlet cubic couplings.

Although in most of the parameter space regions the first-order phase transition is not
strong enough, we demonstrated that there are regions of the parameter space
where a sizeable cubic coupling for a singlet can yield a stochastic
GW signal with peak frequency of $10^{-4}$ to $10^{-2}$~Hz
which may be detectable by future satellites BBO and DECIGO as illustrated
in figure~\ref{fig:GW_coll-BM}.
We also established that such couplings also tend to increase the direct-detection signal,
and may be observable in future detectors such as the XENONnT experiment. The
combination of cubics which suppresses the direct-detection cross
section does not yield a strong enough GW signal.

\section*{Acknowledgements}

This work was supported by the Estonian Research Council grant PRG434, the
grant IUT23-6 of the Estonian Ministry of Education and Research, by the
European Regional Development Fund and programme Mobilitas Pluss grant MOBTT5,
by the European Union through the ERDF Centre of Excellence program project
TK133, and by the Academy of Finland projects
No. 320123 and No. 310130.

\appendix

\section{Shift of the singlet field}
\label{sec:shift}
\subsection{Shifts of parameters under the shift of the field}
\label{sec:shift:param}

We can shift the real part of the complex singlet as $s \to s + \sigma$, e.g., to remove the linear term in the potential. The scalar quartic couplings are shift-invariant, but the dimensionful couplings are shifted as
\begin{align}
  \mu_{1}^{3} &\to \mu_{1}^{3} + (\mu^{2}_{S} + \mu^{\prime 2}_{S}) \sigma
  + \frac{3}{2 \sqrt{2}} (\mu_{3} + \mu'_{3}) \sigma + (\lambda_{S} + \lambda'_{S} + \lambda''_{S}) \sigma^{3},
  \\
  \mu_{3} &\to \mu_{3} + \frac{1}{\sqrt{2}} (4 \lambda'_{S} + \lambda''_{S}) \sigma,
  \\
  \mu'_{3} &\to \mu'_{3} + \frac{1}{\sqrt{2}} (4 \lambda_{S} + 3 \lambda''_{S}) \sigma,
  \\
  \mu^{2}_{H} &\to   \mu^{2}_{H} + \frac{1}{2} \mu_{HS} \sigma + \frac{1}{2} (\lambda_{HS} + \lambda'_{HS}) \sigma^{2},
  \\
  \mu^{2}_{S} &\to \mu^{2}_{S} + \sqrt{2} \mu'_{3} \sigma + \left(2 \lambda_{S} + \frac{3}{2} \lambda''_{S} \right) \sigma^{2},
  \\
  \mu^{\prime 2}_{S} &\to \mu^{\prime 2}_{S} + \frac{1}{\sqrt{2}} (3 \mu_{3} + \mu'_{3}) \sigma
  + \left(\lambda_{S} + 3 \lambda'_{S} + \frac{3}{2} \lambda''_{S} \right) \sigma^{2}.
\end{align}

\subsection{Potential in the shift-invariant notation}

In ref. \cite{Espinosa:2011ax}, the potential of the Higgs boson, $h$, and a real singlet, $s$, is given in terms of shift-invariant quantities. We can extend their formalism to write the potential in eq.~\eqref{original_potential} of $h$, $s$ and $\chi$ in a shift-invariant way as
\begin{equation}
\begin{split}
  V &= \frac{m_{h}^{2}}{8 v^{2}} (h^{2} - v^{2})^{2}
  + \left[ \frac{m_{sh}^{2}}{2 v} (h^{2} - v^{2})
  + \frac{1}{2} (\lambda_{HI} w + 2 \mu_{HS}) \right] (s - w)
  \\
  &+ \frac{1}{4} [\lambda_{HR} (h^{2} - v^{2}) + 2 m_{s}^{2}
  + \lambda_{HI} \chi^{2}] (s - w)^{2},
\end{split}
\end{equation}
where $\lambda_{HR}$ and $\lambda_{HI}$ are defined in eq.~\eqref{eq:vacuum:stability} and $m_{h}^{2}$, $m_{s}^{2}$ and $m_{sh}^{2}$ are the elements of the CP-even scalar mass matrix. A cosmological constant term appearing with a shift has been omitted.

\section{Annihilation cross sections}
\label{sec:x:sections}

Here we define the tree-level annihilation cross sections of the pseudo-Goldstone DM, $\chi$. We use short-hand notations $s_x\equiv \sin x$, and $c_x\equiv \cos x$ for simplicity. It is also useful to define:
\be
\beta_i=\sqrt{1-\frac{4m_i^2}{s}}, \quad
\beta_{ij}=\sqrt{1-\frac{2(m_i^2+m_j^2)}{s}+\frac{(m_i^2-m_j^2)^2}{s^2}},
\quad
k_i=\frac{m_i^2}{s},
\ee
and $v_{\rm rel}=2\beta_\chi$.

The annihilation cross section to fermionic final states is
\be
\sigma_{\chi\chi\to \bar{f}f}=\frac{N_c \beta_{f}^3 m_f^2}{\pi v_{\rm rel} v^2} \left[\frac{c_\theta \lambda_{h_1\chi\chi}}{s-m_1^2}+\frac{s_\theta \lambda_{h_2\chi\chi}}{s-m_2^2}\right]^2,
\ee
where the $N_c$ is $3$ for quarks and $1$ for leptons.

The annihilation cross sections to gauge boson final states are
\begin{align}
\sigma_{\chi\chi\to W^+ W^-}&=\frac{\beta_W}{2\pi v_{\rm rel} s v^2}\left[s^2-4m_W^2 s+12 m_W^4\right]\left[\frac{c_\theta \lambda_{h_1\chi\chi}}{s-m_1^2}+\frac{s_\theta \lambda_{h_2\chi\chi}}{s-m_2^2}\right]^2,\\
\sigma_{\chi\chi\to ZZ}&=\frac{\beta_Z}{2\pi v_{\rm rel} s v^2}\left[s^2-4m_Z^2 s+12 m_Z^4\right]\left[\frac{c_\theta \lambda_{h_1\chi\chi}}{s-m_1^2}+\frac{s_\theta \lambda_{h_2\chi\chi}}{s-m_2^2}\right]^2.
\end{align}

The annihilation cross sections to scalar final states are
\begin{align}
&\sigma_{\chi\chi\to h_i h_i}
=\frac{1}{4\pi}\frac{1}{v_{\rm rel}}\frac{\beta_{i}}{s}\Bigg\{ \alpha^2_i
+ \alpha_i \frac{8\lambda_{h_i\chi\chi}^2 }{s\beta_{\chi}\beta_{i}}\log\left(\frac{1-2k_{i}+\beta_{\chi}\beta_{i}}{1-2k_{i}-\beta_{\chi}\beta_{i}}\right)
\notag
\\
& + \frac{16\lambda_{h_i\chi\chi}^4}{s^2}\left[-\frac{2}{\beta_\chi^2\beta^2_{i}-(1-2k_{i})^2}
+\frac{1}{\beta_\chi\beta_{i}(1-2k_{i})}\log\left(\frac{1-2k_{i}+\beta_{\chi}\beta_{i}}{1-2k_{i}-\beta_{\chi}\beta_{i}}\right)\right]\Bigg\} ,
\\
& \sigma_{\chi\chi\to h_1 h_2}
=\frac{\beta_{12}}{4\pi v_{\rm rel}s}\Bigg\{\alpha_{12}^2+\alpha_{12}\frac{16\lambda_{h_1\chi\chi}\lambda_{h_2\chi\chi}}{s\beta_\chi\beta_{12}}\log\left(\frac{1-k_{1}-k_{2}+\beta_{\chi}\beta_{12}}{1-k_{1}-k_{2}-\beta_{\chi}\beta_{12}}\right)
\notag
\\
&+\frac{32\lambda_{h_1\chi\chi}^2\lambda_{h_2\chi\chi}^2}{s^2}\left[-\frac{2}{\beta_\chi^2\beta_{12}^2-(1-k_1-k_2)^2}
\right.
\notag
\\
&+ \left. \frac{1}{\beta_\chi\beta_{12}(1-k_1-k_2)}\log\left(\frac{1-k_{1}-k_{2}+\beta_{\chi}\beta_{12}}{1-k_{1}-k_{2}-\beta_{\chi}\beta_{12}}\right)\right]\Bigg\},
\end{align}
where
\begin{align}
&\alpha_1=-2\lambda_{h_1 h_1\chi\chi}-\frac{6\lambda_{h_1\chi\chi}\lambda_{h_1 h_1 h_1}}{s-m_1^2}
-\frac{2\lambda_{h_2\chi\chi}\lambda_{h_2 h_1 h_1}}{s-m_2^2},\\
&\alpha_2=-2\lambda_{h_2 h_2\chi\chi}-\frac{2\lambda_{h_1\chi\chi}\lambda_{h_1 h_2 h_2}}{s-m_1^2}
-\frac{6\lambda_{h_2\chi\chi}\lambda_{h_2 h_2 h_2}}{s-m_2^2},\\
&\alpha_{12}=-2\lambda_{h_1 h_2\chi\chi}-\frac{2\lambda_{h_1\chi\chi}\lambda_{h_2 h_1 h_1}}{s-m_1^2}
-\frac{2\lambda_{h_2\chi\chi}\lambda_{h_1 h_2 h_2}}{s-m_2^2}.
\end{align}

The scalar couplings in the above formulae are defined as
\begin{align}
\lambda_{h_1 h_1\chi\chi} =& -\frac{1}{16 v w^3}\Bigg[ 2v w^2 (4w\lambda'_{HS}+\mu_{HS})c^2_\theta
+
4(m_1^2-m_2^2)w^2c^3_\theta s_\theta
\notag
\\
&\qquad\quad+v\Big(-2(m_1^2  +m_2^2 )w+32 w^3 \lambda'_S-4\mu_1^3+3\sqrt{2}w^2(\mu_3+\mu'_3)\\
&\qquad\qquad\quad\ -v^2\mu_{HS}+2(m_1^2-m_2^2)wc_{2\theta}\Big)s^2_\theta\Bigg]
\notag
\\
\lambda_{h_2 h_2\chi\chi} =& -\frac{1}{16 v w^3}\Bigg[ 2v w^2 (4w\lambda'_{HS}+\mu_{HS})s^2_\theta
+
4(m_1^2-m_2^2)w^2c_\theta s^3_\theta
\notag
\\
&\qquad\quad+v\Big(-2(m_1^2  +m_2^2 )w+32 w^3 \lambda'_S-4\mu_1^3+3\sqrt{2}w^2(\mu_3+\mu'_3)\\
&\qquad\qquad\quad\ -v^2\mu_{HS}+2(m_1^2-m_2^2)wc_{2\theta}\Big)c^2_\theta\Bigg]
\notag
\\
\lambda_{h_1 h_2\chi\chi}=&-\frac{c_\theta s_\theta}{8v w^3}\Bigg[
-2(m_1^2-m_2^2)v wc_{2\theta}+2(m_1^2-m_2^2)w^2s_{2\theta}
\notag
\\
&\qquad\quad +v\Big(2(m_1^2+m_2^2)w+8w^3\lambda'_{HS}-32 w^3 \lambda'_{S}+4\mu_1^3\\
&\qquad\qquad\quad -3\sqrt{2}w^2(\mu_3+\mu'_3)+v^2 \mu_{HS} +2w^2\mu_{HS}\Big)\Bigg]
\notag
\\
\lambda_{h_1 h_1 h_1}=& -\frac{1}{16 v w^2}\Bigg[ w(-6m_1^2 w+v^2 \mu_{HS})c_\theta-w(2m_1^2+v^2\mu_{HS})c_{3\theta}
\notag
\\
&\qquad\quad+2v \Big(4m_1^2 w+4\mu_1^3-\sqrt{2}w^2(\mu_3+\mu'_s)+v^2\mu_{HS}\Big)s^3_\theta\Bigg]
\\
\lambda_{h_2 h_2 h_2} =& \frac{1}{32 v w^2}\Bigg[ 3v \Big(4m_2^2 w +4\mu_1^3-\sqrt{2}w^2(\mu_3+\mu'_3)+v^2\mu_{HS}\Big)c_\theta
\notag
\\
&\qquad\qquad\ +v\Big(4m_2^2 w+4\mu_1^3-\sqrt{2}w^2(\mu_3+\mu'_3)+v^2\mu_{HS}\Big) c_{3\theta}
\\
&\qquad\qquad\ -4w \Big(-2m_2^2 w+v^2 \mu_{HS}+(2m_2^2 w+v^2\mu_{HS})c_{2\theta}\Big)s_\theta\Bigg]
\notag
\\
\lambda_{h_1 h_1 h_2} =& \frac{s_\theta}{16 v w^2}\Bigg[
2w(4m_1^2 w +2m_2^2 w +v^2\mu_{HS})
\notag
\\
&\qquad\qquad\ +2w (4m_1^2 w + 2m_2^2 w +3 v^2\mu_{HS})c_{2\theta}
\\
&\qquad\qquad\ +v\Big(8m_1^2 w+4m_2^2 w +12\mu_1^3-3\sqrt{2}w^2(\mu_3+\mu'_3)+3v^2\mu_{HS}\Big)
s_{2\theta}\Bigg],
\notag
\\
\lambda_{h_1 h_2 h_2} =& -\frac{c_\theta}{16 v w^2}\Bigg[-2w(4m_2^2 w +2m_1^2 w +v^2\mu_{HS})+2w (4m_2^2 w + 2m_1^2 w +3 v^2\mu_{HS})c_{2\theta}
\notag
\\
&\qquad\quad\ \ +v\Big(8m_2^2 w+4m_1^2 w +12\mu_1^3-3\sqrt{2}w^2(\mu_3+\mu'_3)+3v^2\mu_{HS}\Big)s_{2\theta}\Bigg].
\end{align}

\section{Direct-detection cross section at one loop}
\label{app:DD1loop}

In specific parts of the parameter space,
the contributions from the symmetry-breaking interactions are suppressed
at tree level. In such case the effect of loop corrections becomes
relevant. Therefore, we will briefly discuss how to extend the analysis
of ref.~\cite{Azevedo:2018exj} to the case of cubic terms.

Let us first briefly summarize the analysis of ref.~\cite{Azevedo:2018exj}
in the U(1)-invariant model with the $\mathbb{Z}_2$-symmetric
mass term. We write the one-loop
contributions to the direct-detection cross section at $t\to 0$ limit as
\begin{equation}
  \label{eq:DD1loop}
  \sigma^{\rm 1L}_{\rm SI} = \frac{f_N^2m_N^2}{4\pi m_{\chi}^2}\left(\frac{m_\chi m_N}{m_\chi+m_N}\right)^2
    |\lambda_{\rm SI}^{\rm 1L}|^2
\end{equation}
In the absence of other symmetry-breaking operators than the mass term for $\chi$,
$\lambda_{\rm SI}^{\rm 1L}$ can be written as
\begin{align}
  \begin{split}
  \label{eq:F0}
  \lambda_{\rm SI,\, 0}^{\rm 1L}=-\frac{s_{2\theta}(m_1^2-m_2^2)m_{\chi}^2}{
      4v^2w^3m_1^2m_2^2}&\left[
      \mathcal{A}_1C_2(0,m_{\chi}^2,m_{\chi}^2,m_1^2,m_2^2,m_{\chi}^2)\right.\\
    &+\mathcal{A}_2D_3(0,0,m_{\chi}^2,m_{\chi}^2,0,m_{\chi}^2,m_1^2,m_1^2,
      m_2^2,m_{\chi}^2)\\
    &\left.+\mathcal{A}_3D_3(0,0,m_{\chi}^2,m_{\chi}^2,0,m_{\chi}^2,m_1^2,m_2^2,
      m_2^2,m_{\chi}^2)
    \right],
  \end{split}
\end{align}
with short-hand notations $s_x\equiv \sin x$ and $c_x\equiv \cos x$, and
\begin{equation}
  \label{eq:A123}
  \begin{split}
    \mathcal{A}_1\equiv& 4(m_1^2s_{\theta}^2+m_2^2c_{\theta}^2)(
      2m_1^2vs_{\theta}^2+2m_2^2vc_{\theta}^2-m_1^2ws_{2\theta}
      +m_2^2ws_{2\theta}),\\
    \mathcal{A}_2\equiv& -2m_1^4s_{\theta}\left[(m_1^2+5m_2^2)wc_{\theta}
      -(m_1^2-m_2^2)(wc_{3\theta}+4vs_{\theta}^3)\right],\\
    \mathcal{A}_3\equiv& 2m_2^4c_{\theta}\left[(5m_1^2+m_2^2)ws_{\theta}
      -(m_1^2-m_2^2)(ws_{3\theta}+4vc_{\theta}^3)\right].
  \end{split}
\end{equation}

Let us then consider what happens in the presence of U(1)-breaking
cubic interactions
$\frac{1}{2} \mu_{3} (S^{3}+S^{* 3})
+ \frac{1}{2} \mu_{3}^{\prime} S S^{*} (S + S^{*})$. For
$\mu_3^{\prime}=-9\mu_3$, the tree-level direct-detection
cross section vanishes in the
limit $t\to 0$, so let us write $\mu_3^{\prime}=-9\mu_3+\delta$. Then
\begin{equation}
  \label{eq:F}
  \begin{split}
    \lambda_{\rm SI}^{\rm 1L}=-\frac{s_{2\theta}(m_1^2-m_2^2)}{
        4v^2w^3m_1^2m_2^2}&\left[\vphantom{\frac12}
        m_\chi^2(\mathcal{A}_1+\delta_{\mathcal{A}_1})
        C_2(0,m_{\chi}^2,m_{\chi}^2,m_1^2,m_2^2,m_{\chi}^2)\right.\\
      &+m_\chi^2(\mathcal{A}_2+\delta_{\mathcal{A}_2})
        D_3(0,0,m_{\chi}^2,m_{\chi}^2,0,m_{\chi}^2,m_1^2,m_1^2,
        m_2^2,m_{\chi}^2)\\
      &+m_\chi^2(\mathcal{A}_3+\delta_{\mathcal{A}_3})
        D_3(0,0,m_{\chi}^2,m_{\chi}^2,0,m_{\chi}^2,m_1^2,m_2^2,
        m_2^2,m_{\chi}^2)\\
      &\left.+\mathcal{A}_4B_0(m_{\chi}^2,m_1^2,m_{\chi}^2)+\mathcal{A}_5 A_0(m_1^2)
        +\mathcal{A}_6A_0(m_{\chi}^2)+\mathcal{O}(\delta)\vphantom{\frac12}
      \right],
  \end{split}
\end{equation}

where
\begin{equation}
  \label{eq:deltaA}
  \begin{split}
    \delta_{\mathcal{A}_1}\equiv&-24\sqrt{2}\mu_3(m_1^2c_{2\theta}
      -2m_2^2c_{\theta}^2)vw+\mathcal{{O}(\delta)},\\
    \delta_{\mathcal{A}_2}\equiv&\ 24\sqrt{2}\mu_3m_1^2(m_1^2-m_2^2)vw
      s_{\theta}^4+\mathcal{{O}(\delta)},\\
    \delta_{\mathcal{A}_3}\equiv&-24\sqrt{2}\mu_3m_2^2(m_1^2-m_2^2)vw
      c_{\theta}^4+\mathcal{{O}(\delta)},\\
    \mathcal{A}_4\equiv&-12\sqrt{2}\mu_3m_1^2vw+\mathcal{O}(\delta),\\
    \mathcal{A}_5\equiv&\ 12\sqrt{2}\mu_3vw+\mathcal{O}(\delta),\\
    \mathcal{A}_6\equiv&\ 6\sqrt{2}\mu_3vw+\mathcal{O}(\delta).
  \end{split}
\end{equation}
The functions $A_0$, $B_0$, $C_2$ and $D_3$ are the standard
Passarino-Veltman functions; our definition of these agrees with the ones given in ref.~\cite{Hahn:1998yk}. The last term in eq.~\eqref{eq:F}, $\mathcal{O}(\delta)$, signifies that for $\delta\neq 0$
several additional loop functions appear that cancel out in $\delta=0$ limit.
Note that while the divergent parts of the loop functions of coefficients
$\mathcal{A}_4$ and $\mathcal{A}_5$ cancel, the divergent part of the last term
$\mathcal{A}_6A_{0}(m_\chi^2)$ does not. (Notice, however,
that the divergent part
vanishes in the limit $m_\chi^2\to0$.) A new counter-term of the form
$\frac12 \delta \mu_3 s\chi^2 $ is needed, where
\begin{equation}
  \label{eq:ctmu3}
  \delta \mu_3=3\sqrt{2}\mu_3\frac{A_0^{\rm div}(m_\chi^2)}{w^2}.
\end{equation}

\section{Renormalisation group equations}
\label{sec:rges}

We use the \texttt{SARAH} code \cite{Staub:2013tta} to calculate the RGEs for
the model. For the sake of conciseness, we present here only the one-loop part. Of couplings to fermions, we take into account only the dominant top quark Yukawa coupling of the Higgs boson. The $\beta$-functions for the quartic couplings are given by
\begin{align}
  16 \pi^{2} \beta_{\lambda_{H}} =&  \frac{3}{8} \left( 3 g^{4}
  + 2 g^{2} g^{\prime 2} + g^{\prime 4} \right) - 3 \lambda_{H} (3 g^{2}
  + g^{\prime 2} - 4 y_{t}^{2})
  \notag
  \\  &+24 \lambda_{H}^2+\lambda_{HS}^2
  +\lambda_{HS}^{\prime 2} - 6 y_{t}^{4},
  \\
  16 \pi^{2} \beta_{\lambda_{HS}} =& \left[ 12 \lambda_{H} + 8 \lambda_{S}
  + 4 \lambda_{HS}  - \frac{3}{2} (3 g^{2} + g^{\prime 2})
  + 6 y_{t}^{2} \right] \lambda_{HS} + 4 \lambda_{HS}^{\prime 2}
  + 6 \lambda'_{HS} \lambda''_{S},
  \\
  16 \pi^{2} \beta_{\lambda'_{HS}} =& \left[ 12 \lambda_{H} +4 \lambda_{S}
  +8 \lambda_{HS} +12  \lambda'_{S} - \frac{3}{2} (3 g^{2} + g^{\prime 2})
  + 6 y_{t}^{2} \right] \lambda'_{HS} +6 \lambda_{HS} \lambda''_{S} ,
  \\
  16 \pi^{2} \beta_{\lambda_{S}} =&\, 20 \lambda_{S}^2+36 \lambda_{S}^{\prime 2}
  + \frac{27}{2} \lambda_{S}^{\prime\prime 2} + 2 \lambda_{HS}^2
  +\lambda_{HS}^{\prime 2},
  \\
  16 \pi^{2} \beta_{\lambda'_{S}} =&\, 24 \lambda_{S} \lambda'_{S}
  +\frac{9}{2} \lambda_{S}^{\prime\prime 2} + \lambda_{HS}^{\prime 2},
  \\
  16 \pi^{2} \beta_{\lambda''_{S}} =&\, 4 \lambda_{HS} \lambda'_{HS}
  + 36 (\lambda_{S}+\lambda'_{S}) \lambda''_{S}.
\end{align}

The $\beta$-functions for the cubic couplings are given by
\begin{align}
  16 \pi^{2} \beta_{\mu_{3}} =&\, 12 \lambda_{S} \mu_{3} + 6 (\lambda'_{S} +
 2 \lambda''_{S}) \mu'_{3} + \sqrt{2} \lambda'_{HS} \mu_{HS},
  \\
  16 \pi^{2} \beta_{\mu'_{3}} =&\, 36 \lambda'_{S} \mu_{3} + 20 \lambda_{S} \mu'_{3} +
 6 \lambda''_{S} (3 \mu_{3} + 2 \mu'_{3}) +
 \sqrt{2} (2  \lambda_{HS} + \lambda'_{HS}) \mu_{HS},
  \\
  16 \pi^{2} \beta_{\mu_{HS}} =& -\frac{3}{4}(g^{\prime 2} + 3 g^2) \mu_{HS} + 3 y_{t}^2 \mu_{HS} +
 6 \lambda_{H} \mu_{HS} +
 2 \lambda_{HS} (\sqrt{2} \mu'_{3} + \mu_{HS})
 \notag
 \\
 &+ \lambda'_{HS} (3 \sqrt{2} \mu_{3} + \sqrt{2} \mu'_{3} + 2 \mu_{HS}).
\end{align}

\bibliography{refs}

\providecommand{\href}[2]{#2}\begingroup\raggedright\begin{thebibliography}{10}

\bibitem{Akerib:2016vxi}
{\scshape LUX} collaboration, D.~S. Akerib et~al., \emph{{Results from a search
  for dark matter in the complete LUX exposure}},
  \href{https://doi.org/10.1103/PhysRevLett.118.021303}{\emph{Phys. Rev. Lett.}
  {\bfseries 118} (2017) 021303}
  [\href{https://arxiv.org/abs/1608.07648}{{\ttfamily 1608.07648}}].

\bibitem{Aprile:2018dbl}
{\scshape XENON} collaboration, E.~Aprile et~al., \emph{{Dark Matter Search
  Results from a One Ton-Year Exposure of XENON1T}},
  \href{https://doi.org/10.1103/PhysRevLett.121.111302}{\emph{Phys. Rev. Lett.}
  {\bfseries 121} (2018) 111302}
  [\href{https://arxiv.org/abs/1805.12562}{{\ttfamily 1805.12562}}].

\bibitem{Cui:2017nnn}
{\scshape PandaX-II} collaboration, X.~Cui et~al., \emph{{Dark Matter Results
  From 54-Ton-Day Exposure of PandaX-II Experiment}},
  \href{https://doi.org/10.1103/PhysRevLett.119.181302}{\emph{Phys. Rev. Lett.}
  {\bfseries 119} (2017) 181302}
  [\href{https://arxiv.org/abs/1708.06917}{{\ttfamily 1708.06917}}].

\bibitem{Gross:2017dan}
C.~Gross, O.~Lebedev and T.~Toma, \emph{{Cancellation Mechanism for
  Dark-Matter–Nucleon Interaction}},
  \href{https://doi.org/10.1103/PhysRevLett.119.191801}{\emph{Phys. Rev. Lett.}
  {\bfseries 119} (2017) 191801}
  [\href{https://arxiv.org/abs/1708.02253}{{\ttfamily 1708.02253}}].

\bibitem{Huitu:2018gbc}
K.~Huitu, N.~Koivunen, O.~Lebedev, S.~Mondal and T.~Toma, \emph{{Probing
  pseudo-Goldstone dark matter at the LHC}},
  \href{https://doi.org/10.1103/PhysRevD.100.015009}{\emph{Phys. Rev.}
  {\bfseries D100} (2019) 015009}
  [\href{https://arxiv.org/abs/1812.05952}{{\ttfamily 1812.05952}}].

\bibitem{Alanne:2018zjm}
T.~Alanne, M.~Heikinheimo, V.~Keus, N.~Koivunen and K.~Tuominen, \emph{{Direct
  and indirect probes of Goldstone dark matter}},
  \href{https://doi.org/10.1103/PhysRevD.99.075028}{\emph{Phys. Rev.}
  {\bfseries D99} (2019) 075028}
  [\href{https://arxiv.org/abs/1812.05996}{{\ttfamily 1812.05996}}].

\bibitem{Azevedo:2018oxv}
D.~Azevedo, M.~Duch, B.~Grzadkowski, D.~Huang, M.~Iglicki and R.~Santos,
  \emph{{Testing scalar versus vector dark matter}},
  \href{https://doi.org/10.1103/PhysRevD.99.015017}{\emph{Phys. Rev.}
  {\bfseries D99} (2019) 015017}
  [\href{https://arxiv.org/abs/1808.01598}{{\ttfamily 1808.01598}}].

\bibitem{Karamitros:2019ewv}
D.~Karamitros, \emph{{Pseudo Nambu-Goldstone Dark Matter: Examples of Vanishing
  Direct Detection Cross Section}},
  \href{https://doi.org/10.1103/PhysRevD.99.095036}{\emph{Phys. Rev. D}
  {\bfseries 99} (2019) 095036}
  [\href{https://arxiv.org/abs/1901.09751}{{\ttfamily 1901.09751}}].

\bibitem{Cline:2019okt}
J.~M. Cline and T.~Toma, \emph{{Pseudo-Goldstone dark matter confronts cosmic
  ray and collider anomalies}},
  \href{https://doi.org/10.1103/PhysRevD.100.035023}{\emph{Phys. Rev.}
  {\bfseries D100} (2019) 035023}
  [\href{https://arxiv.org/abs/1906.02175}{{\ttfamily 1906.02175}}].

\bibitem{Arina:2019tib}
C.~Arina, A.~Beniwal, C.~Degrande, J.~Heisig and A.~Scaffidi, \emph{{Global fit
  of pseudo-Nambu-Goldstone Dark Matter}},
  \href{https://doi.org/10.1007/JHEP04(2020)015}{\emph{JHEP} {\bfseries 04}
  (2020) 015} [\href{https://arxiv.org/abs/1912.04008}{{\ttfamily
  1912.04008}}].

\bibitem{Barger:2008jx}
V.~Barger, P.~Langacker, M.~McCaskey, M.~Ramsey-Musolf and G.~Shaughnessy,
  \emph{{Complex Singlet Extension of the Standard Model}},
  \href{https://doi.org/10.1103/PhysRevD.79.015018}{\emph{Phys. Rev.}
  {\bfseries D79} (2009) 015018}
  [\href{https://arxiv.org/abs/0811.0393}{{\ttfamily 0811.0393}}].

\bibitem{Chiang:2017nmu}
C.-W. Chiang, M.~J. Ramsey-Musolf and E.~Senaha, \emph{{Standard Model with a
  Complex Scalar Singlet: Cosmological Implications and Theoretical
  Considerations}},
  \href{https://doi.org/10.1103/PhysRevD.97.015005}{\emph{Phys. Rev.}
  {\bfseries D97} (2018) 015005}
  [\href{https://arxiv.org/abs/1707.09960}{{\ttfamily 1707.09960}}].

\bibitem{Kannike:2019wsn}
K.~Kannike and M.~Raidal, \emph{{Phase Transitions and Gravitational Wave Tests
  of Pseudo-Goldstone Dark Matter in the Softly Broken U(1) Scalar Singlet
  Model}}, \href{https://doi.org/10.1103/PhysRevD.99.115010}{\emph{Phys. Rev.}
  {\bfseries D99} (2019) 115010}
  [\href{https://arxiv.org/abs/1901.03333}{{\ttfamily 1901.03333}}].

\bibitem{Kannike:2019mzk}
K.~Kannike, K.~Loos and M.~Raidal, \emph{{Gravitational wave signals of
  pseudo-Goldstone dark matter in the $\mathbb{Z}_{3}$ complex singlet model}},
  \href{https://doi.org/10.1103/PhysRevD.101.035001}{\emph{Phys. Rev.}
  {\bfseries D101} (2020) 035001}
  [\href{https://arxiv.org/abs/1907.13136}{{\ttfamily 1907.13136}}].

\bibitem{Witten:1984rs}
E.~Witten, \emph{{Cosmic Separation of Phases}},
  \href{https://doi.org/10.1103/PhysRevD.30.272}{\emph{Phys. Rev.} {\bfseries
  D30} (1984) 272}.

\bibitem{Hogan:1984hx}
C.~J. Hogan, \emph{{NUCLEATION OF COSMOLOGICAL PHASE TRANSITIONS}},
  \href{https://doi.org/10.1016/0370-2693(83)90553-1}{\emph{Phys. Lett.}
  {\bfseries 133B} (1983) 172}.

\bibitem{Steinhardt:1981ct}
P.~J. Steinhardt, \emph{{Relativistic Detonation Waves and Bubble Growth in
  False Vacuum Decay}},
  \href{https://doi.org/10.1103/PhysRevD.25.2074}{\emph{Phys. Rev.} {\bfseries
  D25} (1982) 2074}.

\bibitem{Audley:2017drz}
{\scshape LISA} collaboration, P.~Amaro-Seoane et~al., \emph{{Laser
  Interferometer Space Antenna}},
  \href{https://arxiv.org/abs/1702.00786}{{\ttfamily 1702.00786}}.

\bibitem{Baker:2019nia}
J.~Baker et~al., \emph{{The Laser Interferometer Space Antenna: Unveiling the
  Millihertz Gravitational Wave Sky}},
  \href{https://arxiv.org/abs/1907.06482}{{\ttfamily 1907.06482}}.

\bibitem{Seto:2001qf}
N.~Seto, S.~Kawamura and T.~Nakamura, \emph{{Possibility of direct measurement
  of the acceleration of the universe using 0.1-Hz band laser interferometer
  gravitational wave antenna in space}},
  \href{https://doi.org/10.1103/PhysRevLett.87.221103}{\emph{Phys. Rev. Lett.}
  {\bfseries 87} (2001) 221103}
  [\href{https://arxiv.org/abs/astro-ph/0108011}{{\ttfamily
  astro-ph/0108011}}].

\bibitem{Kawamura:2006up}
S.~Kawamura et~al., \emph{{The Japanese space gravitational wave antenna
  DECIGO}}, \href{https://doi.org/10.1088/0264-9381/23/8/S17}{\emph{Class.
  Quant. Grav.} {\bfseries 23} (2006) S125}.

\bibitem{Yagi:2011wg}
K.~Yagi and N.~Seto, \emph{{Detector configuration of DECIGO/BBO and
  identification of cosmological neutron-star binaries}},
  \href{https://doi.org/10.1103/PhysRevD.95.109901,
  10.1103/PhysRevD.83.044011}{\emph{Phys. Rev.} {\bfseries D83} (2011) 044011}
  [\href{https://arxiv.org/abs/1101.3940}{{\ttfamily 1101.3940}}].

\bibitem{Isoyama:2018rjb}
S.~Isoyama, H.~Nakano and T.~Nakamura, \emph{{Multiband Gravitational-Wave
  Astronomy: Observing binary inspirals with a decihertz detector, B-DECIGO}},
  \href{https://doi.org/10.1093/ptep/pty078}{\emph{PTEP} {\bfseries 2018}
  (2018) 073E01} [\href{https://arxiv.org/abs/1802.06977}{{\ttfamily
  1802.06977}}].

\bibitem{Crowder:2005nr}
J.~Crowder and N.~J. Cornish, \emph{{Beyond LISA: Exploring future
  gravitational wave missions}},
  \href{https://doi.org/10.1103/PhysRevD.72.083005}{\emph{Phys. Rev.}
  {\bfseries D72} (2005) 083005}
  [\href{https://arxiv.org/abs/gr-qc/0506015}{{\ttfamily gr-qc/0506015}}].

\bibitem{Corbin:2005ny}
V.~Corbin and N.~J. Cornish, \emph{{Detecting the cosmic gravitational wave
  background with the big bang observer}},
  \href{https://doi.org/10.1088/0264-9381/23/7/014}{\emph{Class. Quant. Grav.}
  {\bfseries 23} (2006) 2435}
  [\href{https://arxiv.org/abs/gr-qc/0512039}{{\ttfamily gr-qc/0512039}}].

\bibitem{Harry:2006fi}
G.~M. Harry, P.~Fritschel, D.~A. Shaddock, W.~Folkner and E.~S. Phinney,
  \emph{{Laser interferometry for the big bang observer}},
  \href{https://doi.org/10.1088/0264-9381/23/24/C01,
  10.1088/0264-9381/23/15/008}{\emph{Class. Quant. Grav.} {\bfseries 23} (2006)
  4887}.

\bibitem{Jiang:2015cwa}
M.~Jiang, L.~Bian, W.~Huang and J.~Shu, \emph{{Impact of a complex singlet:
  Electroweak baryogenesis and dark matter}},
  \href{https://doi.org/10.1103/PhysRevD.93.065032}{\emph{Phys. Rev.}
  {\bfseries D93} (2016) 065032}
  [\href{https://arxiv.org/abs/1502.07574}{{\ttfamily 1502.07574}}].

\bibitem{Alves:2018oct}
A.~Alves, T.~Ghosh, H.-K. Guo and K.~Sinha, \emph{{Resonant Di-Higgs Production
  at Gravitational Wave Benchmarks: A Collider Study using Machine Learning}},
  \href{https://doi.org/10.1007/JHEP12(2018)070}{\emph{JHEP} {\bfseries 12}
  (2018) 070} [\href{https://arxiv.org/abs/1808.08974}{{\ttfamily
  1808.08974}}].

\bibitem{Alves:2018jsw}
A.~Alves, T.~Ghosh, H.-K. Guo, K.~Sinha and D.~Vagie, \emph{{Collider and
  Gravitational Wave Complementarity in Exploring the Singlet Extension of the
  Standard Model}}, \href{https://doi.org/10.1007/JHEP04(2019)052}{\emph{JHEP}
  {\bfseries 04} (2019) 052}
  [\href{https://arxiv.org/abs/1812.09333}{{\ttfamily 1812.09333}}].

\bibitem{Azevedo:2018exj}
D.~Azevedo, M.~Duch, B.~Grzadkowski, D.~Huang, M.~Iglicki and R.~Santos,
  \emph{{One-loop contribution to dark-matter-nucleon scattering in the
  pseudo-scalar dark matter model}},
  \href{https://doi.org/10.1007/JHEP01(2019)138}{\emph{JHEP} {\bfseries 01}
  (2019) 138} [\href{https://arxiv.org/abs/1810.06105}{{\ttfamily
  1810.06105}}].

\bibitem{Ishiwata:2018sdi}
K.~Ishiwata and T.~Toma, \emph{{Probing pseudo Nambu-Goldstone boson dark
  matter at loop level}},
  \href{https://doi.org/10.1007/JHEP12(2018)089}{\emph{JHEP} {\bfseries 12}
  (2018) 089} [\href{https://arxiv.org/abs/1810.08139}{{\ttfamily
  1810.08139}}].

\bibitem{Chao:2017vrq}
W.~Chao, H.-K. Guo and J.~Shu, \emph{{Gravitational Wave Signals of Electroweak
  Phase Transition Triggered by Dark Matter}},
  \href{https://doi.org/10.1088/1475-7516/2017/09/009}{\emph{JCAP} {\bfseries
  09} (2017) 009} [\href{https://arxiv.org/abs/1702.02698}{{\ttfamily
  1702.02698}}].

\bibitem{Branco:1999fs}
G.~C. Branco, L.~Lavoura and J.~P. Silva, \emph{{CP Violation}}, {\emph{Int.
  Ser. Monogr. Phys.} {\bfseries 103} (1999) 1}.

\bibitem{Kannike:2012pe}
K.~Kannike, \emph{{Vacuum Stability Conditions From Copositivity Criteria}},
  \href{https://doi.org/10.1140/epjc/s10052-012-2093-z}{\emph{Eur. Phys. J.}
  {\bfseries C72} (2012) 2093}
  [\href{https://arxiv.org/abs/1205.3781}{{\ttfamily 1205.3781}}].

\bibitem{Kannike:2016fmd}
K.~Kannike, \emph{{Vacuum Stability of a General Scalar Potential of a Few
  Fields}}, \href{https://doi.org/10.1140/epjc/s10052-016-4160-3,
  10.1140/epjc/s10052-018-5837-6}{\emph{Eur. Phys. J.} {\bfseries C76} (2016)
  324} [\href{https://arxiv.org/abs/1603.02680}{{\ttfamily 1603.02680}}].

\bibitem{Kanemura:1993hm}
S.~Kanemura, T.~Kubota and E.~Takasugi, \emph{{Lee-Quigg-Thacker bounds for
  Higgs boson masses in a two doublet model}},
  \href{https://doi.org/10.1016/0370-2693(93)91205-2}{\emph{Phys. Lett.}
  {\bfseries B313} (1993) 155}
  [\href{https://arxiv.org/abs/hep-ph/9303263}{{\ttfamily hep-ph/9303263}}].

\bibitem{Akeroyd:2000wc}
A.~G. Akeroyd, A.~Arhrib and E.-M. Naimi, \emph{{Note on tree level unitarity
  in the general two Higgs doublet model}},
  \href{https://doi.org/10.1016/S0370-2693(00)00962-X}{\emph{Phys. Lett.}
  {\bfseries B490} (2000) 119}
  [\href{https://arxiv.org/abs/hep-ph/0006035}{{\ttfamily hep-ph/0006035}}].

\bibitem{Goodsell:2018tti}
M.~D. Goodsell and F.~Staub, \emph{{Unitarity constraints on general scalar
  couplings with SARAH}},
  \href{https://doi.org/10.1140/epjc/s10052-018-6127-z}{\emph{Eur. Phys. J.}
  {\bfseries C78} (2018) 649}
  [\href{https://arxiv.org/abs/1805.07306}{{\ttfamily 1805.07306}}].

\bibitem{10.2307/2285901}
P.~A. Samuelson, \emph{How deviant can you be?}, {\emph{Journal of the American
  Statistical Association} {\bfseries 63} (1968) 1522}.

\bibitem{Belanger:2014bga}
G.~Bélanger, K.~Kannike, A.~Pukhov and M.~Raidal, \emph{{Minimal
  semi-annihilating $\mathbb{Z}_N$ scalar dark matter}},
  \href{https://doi.org/10.1088/1475-7516/2014/06/021}{\emph{JCAP} {\bfseries
  1406} (2014) 021} [\href{https://arxiv.org/abs/1403.4960}{{\ttfamily
  1403.4960}}].

\bibitem{Khachatryan:2016vau}
{\scshape ATLAS, CMS} collaboration, G.~Aad et~al., \emph{{Measurements of the
  Higgs boson production and decay rates and constraints on its couplings from
  a combined ATLAS and CMS analysis of the LHC pp collision data at $
  \sqrt{s}=7 $ and 8 TeV}},
  \href{https://doi.org/10.1007/JHEP08(2016)045}{\emph{JHEP} {\bfseries 08}
  (2016) 045} [\href{https://arxiv.org/abs/1606.02266}{{\ttfamily
  1606.02266}}].

\bibitem{Beacham:2019nyx}
J.~Beacham et~al., \emph{{Physics Beyond Colliders at CERN: Beyond the Standard
  Model Working Group Report}},
  \href{https://doi.org/10.1088/1361-6471/ab4cd2}{\emph{J. Phys.} {\bfseries
  G47} (2020) 010501} [\href{https://arxiv.org/abs/1901.09966}{{\ttfamily
  1901.09966}}].

\bibitem{Sirunyan:2019twz}
{\scshape CMS} collaboration, A.~M. Sirunyan et~al., \emph{{Measurements of the
  Higgs boson width and anomalous $HVV$ couplings from on-shell and off-shell
  production in the four-lepton final state}},
  \href{https://doi.org/10.1103/PhysRevD.99.112003}{\emph{Phys. Rev.}
  {\bfseries D99} (2019) 112003}
  [\href{https://arxiv.org/abs/1901.00174}{{\ttfamily 1901.00174}}].

\bibitem{Khachatryan:2016whc}
{\scshape CMS} collaboration, V.~Khachatryan et~al., \emph{{Searches for
  invisible decays of the Higgs boson in pp collisions at $\sqrt{s}$ = 7, 8,
  and 13 TeV}}, \href{https://doi.org/10.1007/JHEP02(2017)135}{\emph{JHEP}
  {\bfseries 02} (2017) 135}
  [\href{https://arxiv.org/abs/1610.09218}{{\ttfamily 1610.09218}}].

\bibitem{ATLAS-CONF-2018-031}
{ATLAS collaboration}, \emph{{Combined measurements of Higgs boson production
  and decay using up to 80 fb$^{-1}$ of proton--proton collision data at
  $\sqrt{s}=$ 13 TeV collected with the ATLAS experiment}}, .

\bibitem{Aghanim:2018eyx}
{\scshape Planck} collaboration, N.~Aghanim et~al., \emph{{Planck 2018 results.
  VI. Cosmological parameters}},
  \href{https://arxiv.org/abs/1807.06209}{{\ttfamily 1807.06209}}.

\bibitem{Alarcon:2011zs}
J.~M. Alarcon, J.~Martin~Camalich and J.~A. Oller, \emph{{The chiral
  representation of the $\pi N$ scattering amplitude and the pion-nucleon sigma
  term}}, \href{https://doi.org/10.1103/PhysRevD.85.051503}{\emph{Phys. Rev.}
  {\bfseries D85} (2012) 051503}
  [\href{https://arxiv.org/abs/1110.3797}{{\ttfamily 1110.3797}}].

\bibitem{Alarcon:2012nr}
J.~M. Alarcon, L.~S. Geng, J.~Martin~Camalich and J.~A. Oller, \emph{{The
  strangeness content of the nucleon from effective field theory and
  phenomenology}},
  \href{https://doi.org/10.1016/j.physletb.2014.01.065}{\emph{Phys. Lett.}
  {\bfseries B730} (2014) 342}
  [\href{https://arxiv.org/abs/1209.2870}{{\ttfamily 1209.2870}}].

\bibitem{Cline:2013gha}
J.~M. Cline, K.~Kainulainen, P.~Scott and C.~Weniger, \emph{{Update on scalar
  singlet dark matter}}, \href{https://doi.org/10.1103/PhysRevD.92.039906,
  10.1103/PhysRevD.88.055025}{\emph{Phys. Rev.} {\bfseries D88} (2013) 055025}
  [\href{https://arxiv.org/abs/1306.4710}{{\ttfamily 1306.4710}}].

\bibitem{Borowka:2017idc}
S.~Borowka, G.~Heinrich, S.~Jahn, S.~P. Jones, M.~Kerner, J.~Schlenk et~al.,
  \emph{{pySecDec: a toolbox for the numerical evaluation of multi-scale
  integrals}}, \href{https://doi.org/10.1016/j.cpc.2017.09.015}{\emph{Comput.
  Phys. Commun.} {\bfseries 222} (2018) 313}
  [\href{https://arxiv.org/abs/1703.09692}{{\ttfamily 1703.09692}}].

\bibitem{Borowka:2018goh}
S.~Borowka, G.~Heinrich, S.~Jahn, S.~P. Jones, M.~Kerner and J.~Schlenk,
  \emph{{A GPU compatible quasi-Monte Carlo integrator interfaced to
  pySecDec}}, \href{https://doi.org/10.1016/j.cpc.2019.02.015}{\emph{Comput.
  Phys. Commun.} {\bfseries 240} (2019) 120}
  [\href{https://arxiv.org/abs/1811.11720}{{\ttfamily 1811.11720}}].

\bibitem{Vermaseren:2000nd}
J.~A.~M. Vermaseren, \emph{{New features of FORM}},
  \href{https://arxiv.org/abs/math-ph/0010025}{{\ttfamily math-ph/0010025}}.

\bibitem{Kuipers:2013pba}
J.~Kuipers, T.~Ueda and J.~A.~M. Vermaseren, \emph{{Code Optimization in
  FORM}}, \href{https://doi.org/10.1016/j.cpc.2014.08.008}{\emph{Comput. Phys.
  Commun.} {\bfseries 189} (2015) 1}
  [\href{https://arxiv.org/abs/1310.7007}{{\ttfamily 1310.7007}}].

\bibitem{Ruijl:2017dtg}
B.~Ruijl, T.~Ueda and J.~Vermaseren, \emph{{FORM version 4.2}},
  \href{https://arxiv.org/abs/1707.06453}{{\ttfamily 1707.06453}}.

\bibitem{Hahn:2004fe}
T.~Hahn, \emph{{CUBA: A Library for multidimensional numerical integration}},
  \href{https://doi.org/10.1016/j.cpc.2005.01.010}{\emph{Comput. Phys. Commun.}
  {\bfseries 168} (2005) 78}
  [\href{https://arxiv.org/abs/hep-ph/0404043}{{\ttfamily hep-ph/0404043}}].

\bibitem{Hahn:2014fua}
T.~Hahn, \emph{{Concurrent Cuba}},
  \href{https://doi.org/10.1088/1742-6596/608/1/012066}{\emph{J. Phys. Conf.
  Ser.} {\bfseries 608} (2015) 012066}
  [\href{https://arxiv.org/abs/1408.6373}{{\ttfamily 1408.6373}}].

\bibitem{Boddy:2019qak}
K.~K. Boddy, J.~Kumar, A.~B. Pace, J.~Runburg and L.~E. Strigari,
  \emph{{Effective $J$-factors for Milky Way dwarf spheroidal galaxies with
  velocity-dependent annihilation}},
  \href{https://doi.org/10.1103/PhysRevD.102.023029}{\emph{Phys. Rev.}
  {\bfseries D102} (2020) 023029}
  [\href{https://arxiv.org/abs/1909.13197}{{\ttfamily 1909.13197}}].

\bibitem{Fermi-LAT:2016uux}
{\scshape Fermi-LAT, DES} collaboration, A.~Albert et~al., \emph{{Searching for
  Dark Matter Annihilation in Recently Discovered Milky Way Satellites with
  Fermi-LAT}},
  \href{https://doi.org/10.3847/1538-4357/834/2/110}{\emph{Astrophys. J.}
  {\bfseries 834} (2017) 110}
  [\href{https://arxiv.org/abs/1611.03184}{{\ttfamily 1611.03184}}].

\bibitem{Clark:2017fum}
S.~J. Clark, B.~Dutta and L.~E. Strigari, \emph{{Dark Matter Annihilation into
  Four-Body Final States and Implications for the AMS Antiproton Excess}},
  \href{https://doi.org/10.1103/PhysRevD.97.023003}{\emph{Phys. Rev.}
  {\bfseries D97} (2018) 023003}
  [\href{https://arxiv.org/abs/1709.07410}{{\ttfamily 1709.07410}}].

\bibitem{Boddy:2018qur}
K.~Boddy, J.~Kumar, D.~Marfatia and P.~Sandick, \emph{{Model-independent
  constraints on dark matter annihilation in dwarf spheroidal galaxies}},
  \href{https://doi.org/10.1103/PhysRevD.97.095031}{\emph{Phys. Rev.}
  {\bfseries D97} (2018) 095031}
  [\href{https://arxiv.org/abs/1802.03826}{{\ttfamily 1802.03826}}].

\bibitem{Kainulainen:2019kyp}
K.~Kainulainen, V.~Keus, L.~Niemi, K.~Rummukainen, T.~V.~I. Tenkanen and
  V.~Vaskonen, \emph{{On the validity of perturbative studies of the
  electroweak phase transition in the Two Higgs Doublet model}},
  \href{https://doi.org/10.1007/JHEP06(2019)075}{\emph{JHEP} {\bfseries 06}
  (2019) 075} [\href{https://arxiv.org/abs/1904.01329}{{\ttfamily
  1904.01329}}].

\bibitem{Ellis:2019oqb}
J.~Ellis, M.~Lewicki, J.~M. No and V.~Vaskonen, \emph{{Gravitational wave
  energy budget in strongly supercooled phase transitions}},
  \href{https://doi.org/10.1088/1475-7516/2019/06/024}{\emph{JCAP} {\bfseries
  1906} (2019) 024} [\href{https://arxiv.org/abs/1903.09642}{{\ttfamily
  1903.09642}}].

\bibitem{Grojean:2006bp}
C.~Grojean and G.~Servant, \emph{{Gravitational Waves from Phase Transitions at
  the Electroweak Scale and Beyond}},
  \href{https://doi.org/10.1103/PhysRevD.75.043507}{\emph{Phys. Rev.}
  {\bfseries D75} (2007) 043507}
  [\href{https://arxiv.org/abs/hep-ph/0607107}{{\ttfamily hep-ph/0607107}}].

\bibitem{Caprini:2009yp}
C.~Caprini, R.~Durrer and G.~Servant, \emph{{The stochastic gravitational wave
  background from turbulence and magnetic fields generated by a first-order
  phase transition}},
  \href{https://doi.org/10.1088/1475-7516/2009/12/024}{\emph{JCAP} {\bfseries
  0912} (2009) 024} [\href{https://arxiv.org/abs/0909.0622}{{\ttfamily
  0909.0622}}].

\bibitem{Caprini:2015zlo}
C.~Caprini et~al., \emph{{Science with the space-based interferometer eLISA.
  II: Gravitational waves from cosmological phase transitions}},
  \href{https://doi.org/10.1088/1475-7516/2016/04/001}{\emph{JCAP} {\bfseries
  1604} (2016) 001} [\href{https://arxiv.org/abs/1512.06239}{{\ttfamily
  1512.06239}}].

\bibitem{Axen:2018zvb}
M.~Fitz~Axen, S.~Banagiri, A.~Matas, C.~Caprini and V.~Mandic,
  \emph{{Multiwavelength observations of cosmological phase transitions using
  LISA and Cosmic Explorer}},
  \href{https://doi.org/10.1103/PhysRevD.98.103508}{\emph{Phys. Rev.}
  {\bfseries D98} (2018) 103508}
  [\href{https://arxiv.org/abs/1806.02500}{{\ttfamily 1806.02500}}].

\bibitem{Guo:2020grp}
H.-K. Guo, K.~Sinha, D.~Vagie and G.~White, \emph{{Phase Transitions in an
  Expanding Universe: Stochastic Gravitational Waves in Standard and
  Non-Standard Histories}},  \href{https://arxiv.org/abs/2007.08537}{{\ttfamily
  2007.08537}}.

\bibitem{Allen:1997ad}
B.~Allen and J.~D. Romano, \emph{{Detecting a stochastic background of
  gravitational radiation: Signal processing strategies and sensitivities}},
  \href{https://doi.org/10.1103/PhysRevD.59.102001}{\emph{Phys. Rev.}
  {\bfseries D59} (1999) 102001}
  [\href{https://arxiv.org/abs/gr-qc/9710117}{{\ttfamily gr-qc/9710117}}].

\bibitem{Maggiore:1999vm}
M.~Maggiore, \emph{{Gravitational wave experiments and early universe
  cosmology}}, \href{https://doi.org/10.1016/S0370-1573(99)00102-7}{\emph{Phys.
  Rept.} {\bfseries 331} (2000) 283}
  [\href{https://arxiv.org/abs/gr-qc/9909001}{{\ttfamily gr-qc/9909001}}].

\bibitem{Schmitz:2020syl}
K.~Schmitz, \emph{{New Sensitivity Curves for Gravitational-Wave Experiments}},
   \href{https://arxiv.org/abs/2002.04615}{{\ttfamily 2002.04615}}.

\bibitem{Alanne:2019bsm}
T.~Alanne, T.~Hugle, M.~Platscher and K.~Schmitz, \emph{{A fresh look at the
  gravitational-wave signal from cosmological phase transitions}},
  \href{https://doi.org/10.1007/JHEP03(2020)004}{\emph{JHEP} {\bfseries 03}
  (2020) 004} [\href{https://arxiv.org/abs/1909.11356}{{\ttfamily
  1909.11356}}].

\bibitem{Thrane:2013oya}
E.~Thrane and J.~D. Romano, \emph{{Sensitivity curves for searches for
  gravitational-wave backgrounds}},
  \href{https://doi.org/10.1103/PhysRevD.88.124032}{\emph{Phys. Rev.}
  {\bfseries D88} (2013) 124032}
  [\href{https://arxiv.org/abs/1310.5300}{{\ttfamily 1310.5300}}].

\bibitem{Belanger:2018mqt}
G.~Bélanger, F.~Boudjema, A.~Goudelis, A.~Pukhov and B.~Zaldivar,
  \emph{{micrOMEGAs5.0 : Freeze-in}},
  \href{https://doi.org/10.1016/j.cpc.2018.04.027}{\emph{Comput. Phys. Commun.}
  {\bfseries 231} (2018) 173}
  [\href{https://arxiv.org/abs/1801.03509}{{\ttfamily 1801.03509}}].

\bibitem{Christensen:2008py}
N.~D. Christensen and C.~Duhr, \emph{{FeynRules - Feynman rules made easy}},
  \href{https://doi.org/10.1016/j.cpc.2009.02.018}{\emph{Comput. Phys. Commun.}
  {\bfseries 180} (2009) 1614}
  [\href{https://arxiv.org/abs/0806.4194}{{\ttfamily 0806.4194}}].

\bibitem{Christensen:2009jx}
N.~D. Christensen, P.~de~Aquino, C.~Degrande, C.~Duhr, B.~Fuks, M.~Herquet
  et~al., \emph{{A Comprehensive approach to new physics simulations}},
  \href{https://doi.org/10.1140/epjc/s10052-011-1541-5}{\emph{Eur. Phys. J.}
  {\bfseries C71} (2011) 1541}
  [\href{https://arxiv.org/abs/0906.2474}{{\ttfamily 0906.2474}}].

\bibitem{Alloul:2013bka}
A.~Alloul, N.~D. Christensen, C.~Degrande, C.~Duhr and B.~Fuks,
  \emph{{FeynRules 2.0 - A complete toolbox for tree-level phenomenology}},
  \href{https://doi.org/10.1016/j.cpc.2014.04.012}{\emph{Comput. Phys. Commun.}
  {\bfseries 185} (2014) 2250}
  [\href{https://arxiv.org/abs/1310.1921}{{\ttfamily 1310.1921}}].

\bibitem{Wainwright:2011kj}
C.~L. Wainwright, \emph{{CosmoTransitions: Computing Cosmological Phase
  Transition Temperatures and Bubble Profiles with Multiple Fields}},
  \href{https://doi.org/10.1016/j.cpc.2012.04.004}{\emph{Comput. Phys. Commun.}
  {\bfseries 183} (2012) 2006}
  [\href{https://arxiv.org/abs/1109.4189}{{\ttfamily 1109.4189}}].

\bibitem{Guada:2020xnz}
V.~Guada, M.~Nemevšek and M.~Pintar, \emph{{FindBounce: package for
  multi-field bounce actions}},
  \href{https://doi.org/10.1016/j.cpc.2020.107480}{\emph{Comput. Phys. Commun.}
  {\bfseries 256} (2020) 107480}
  [\href{https://arxiv.org/abs/2002.00881}{{\ttfamily 2002.00881}}].

\bibitem{Espinosa:2011ax}
J.~R. Espinosa, T.~Konstandin and F.~Riva, \emph{{Strong Electroweak Phase
  Transitions in the Standard Model with a Singlet}},
  \href{https://doi.org/10.1016/j.nuclphysb.2011.09.010}{\emph{Nucl. Phys.}
  {\bfseries B854} (2012) 592}
  [\href{https://arxiv.org/abs/1107.5441}{{\ttfamily 1107.5441}}].

\bibitem{Hahn:1998yk}
T.~Hahn and M.~Perez-Victoria, \emph{{Automatized one loop calculations in
  four-dimensions and D-dimensions}},
  \href{https://doi.org/10.1016/S0010-4655(98)00173-8}{\emph{Comput. Phys.
  Commun.} {\bfseries 118} (1999) 153}
  [\href{https://arxiv.org/abs/hep-ph/9807565}{{\ttfamily hep-ph/9807565}}].

\bibitem{Staub:2013tta}
F.~Staub, \emph{{SARAH 4 : A tool for (not only SUSY) model builders}},
  \href{https://doi.org/10.1016/j.cpc.2014.02.018}{\emph{Comput. Phys. Commun.}
  {\bfseries 185} (2014) 1773}
  [\href{https://arxiv.org/abs/1309.7223}{{\ttfamily 1309.7223}}].

\end{thebibliography}\endgroup

\end{document}